\documentclass[12pt,subeqn]{iopart}

\usepackage{iopams}
\usepackage{setstack}
\usepackage{graphicx}
\usepackage{epstopdf}
\begin{document}

\title[Quantum--optical phenomena with optical lattices]
{Simulating quantum--optical phenomena
\\
with cold atoms in optical lattices}

\author{Carlos Navarrete--Benlloch$^1$, In\'es de Vega$^2$, Diego Porras$^3$, and J. Ignacio Cirac$^4$}

\address{$^1$ Departament d'\`Optica, Universitat de Val\`encia, Dr. Moliner 50, 46100 Burjassot, Spain}
\address{$^2$ Institut f\"ur Theoretische Physik, Albert-Einstein-Allee 11, Universit\"at Ulm, D-89069 Ulm, Germany.}
\address{$^3$ Departamento de F\'isica Te\'orica I, Universidad Complutense, 28040 Madrid, Spain}
\address{$^4$ Max-Planck-Institut f\"ur Quantenoptik, Hans-Kopfermann-Strasse 1, 85748 Garching, Germany}
\eads{\mailto{carlos.navarrete@uv.es}, \mailto{ines.devega@uni-ulm.de}, \mailto{diego.porras@fis.ucm.es}, \mailto{ignacio.cirac@mpq.mpg.de}}

\begin{abstract}
We propose a scheme involving cold atoms trapped in optical lattices to observe different phenomena traditionally linked to quantum--optical systems. The basic idea consists of connecting the trapped atomic state to a non-trapped state through a Raman scheme. The coupling between these two types of atoms (trapped and free) turns out to be similar to that describing light--matter interaction within the rotating--wave approximation, the role of matter and photons being played by the trapped and free atoms, respectively. We explain in particular how to observe phenomena arising from the collective spontaneous emission of atomic and harmonic oscillator samples such as superradiance and directional emission. We also show how the same setup can simulate Bose--Hubbard Hamiltonians with extended hopping as well as Ising models with long--range interactions. We believe that this system can be realized with state of the art technology.
\end{abstract}


\maketitle

\section{Introduction \label{Introduction}}

Cold atoms trapped in optical lattices have been proved to be very versatile quantum systems in which a large class of many--body condensed--matter
Hamiltonians can be simulated (see \cite{Jaksch04rev,Lewenstein07rev,Bloch08rev} for extensive reviews on this subject). One of the first proposals in this direction was the article by Jacksch et al. \cite{Jaksch98}, where it was proved that the dynamics of cold atoms trapped in optical lattices is described by the Bose--Hubbard Hamiltonian provided that certain conditions are satisfied; shortly after, the superfluid--to--Mott insulator phase transition characteristic of this Hamiltonian was observed in the laboratory \cite{Greiner02}. Since then, the broad tunability of the lattice parameters, and the increasing ability to trap different kind of particles (like bosonic and fermionic atoms with arbitrary spin or polar molecules), has allowed theoreticians to propose optical lattices as promising simulators for different types of generalized Bose--Hubbard and spin models which are in close relation to important condensed--matter phenomena \cite{Jaksch04rev,Lewenstein07rev,Bloch08rev}. Recent experiments have shown that optical lattices can be used to address open problems in physics like, e.g., high--$T_\mathrm{c}$ superconductivity \cite{Schneider08}, to study phenomena in low dimensions such as the Berezinskii--Kosterlitz--Thouless transition \cite{Hadzibabic06}, or to implement quantum computation schemes \cite{Anderlini07}.

In this article we keep digging into the capabilities of optical lattices as simulators, showing how they can also be a powerful tool for simulating
quantum--optical phenomena. In particular, following our previous proposal \cite{deVega08}, we introduce various schemes in which superradiance-like
phenomena can be observed. In addition, we will show that these same schemes realize Bose--Hubbard and spin models with extended interactions (in particular, models where hopping between sites and ferromagnetic interactions between spins are long range instead of nearest--neighbors).

The concept of superradiance was introduced by Dicke in 1954 when studying the spontaneous emission of a collection of two--level atoms \cite{Dicke54} (see \cite{Gross82rev} for a review). He showed that certain collective states where the excitations are distributed symmetrically over the whole sample have enhanced emission rates. Probably the most stunning example is the single--excitation symmetric state (now known as the symmetric Dicke state), which instead of decaying with the single--atom decay rate $\Gamma_0$, was shown to decay with $N\Gamma_0$, $N$ being the number of atoms. He also suggested that the emission rate of the state having all the atoms excited should be enhanced at the initial steps of the decay process, which was a most interesting prediction from the experimental point of view, as this state is in general easier to prepare. However, Dicke used a very simplified model in which all the atoms interact with a common radiation field within the dipolar approximation; almost 15 years later, and motivated by the new atom--inversion techniques, several authors showed that dipolar interactions impose a threshold value for the atom density, that is, for the number of interacting atoms, in order for superradiance to appear \cite{Ernst68,Agarwal70,Rehler71}.

It is worth noting that together with atomic ensembles, spontaneous emission of collections of harmonic oscillators were also studied at that time
\cite{Agarwal70}. Two interesting features of this system were reported: (i) an initial state with all the harmonic oscillators excited does not show rate enhancement, on the contrary, most excitations remain within the sample in the steady state, while (ii) the state having all the oscillators in the same coherent state has a superradiant rate.

In the last years, there has been a renewed interest in this phenomenology because of its potential for quantum technologies
\cite{Eberly06,Scully06,Porras07a,Porras07,Scully09a,Scully09b}. In these new approaches, superradiant effects differ from that found by Dicke in that they are mediated by the initial entanglement of the emitters, not by a build up of their coherence. Of particular interest to our current work is the analysis performed in \cite{Porras07} (see also \cite{Porras07a}) of the spontaneous emission by regular arrays of atoms separated a distance $d_0$ comparable to the wavelength $\lambda$ of the emitted radiation. It was shown that directional emission can be obtained for particular single--excitation entangled states when $\lambda>2d_0$, the emission rate being enhanced by a factor $\chi\propto\left(\lambda/d_0\right)^2N^{1/3}$.

Also of interest for our purposes is the work carried by Sajeev John and collaborators on the collective emission of atoms embedded in photonic
band--gap materials \cite{John94,John95}. In this kind of materials the density of states of the electromagnetic field is zero for frequencies laying
within the gap, what gives rise to the phenomenon of light localization \cite{John84}. As for collective emission, it was shown by using a simplified
Dicke-like model \cite{John94,John95} that when the atomic transition lies close to the gap, radiation is emitted in the form of evanescent waves and the single--excitation symmetric state has a rate proportional to $N^2\Gamma_0$, which is even larger than the one predicted by Dicke in free space.

\begin{figure}[t]

\flushright \includegraphics[width=5in]{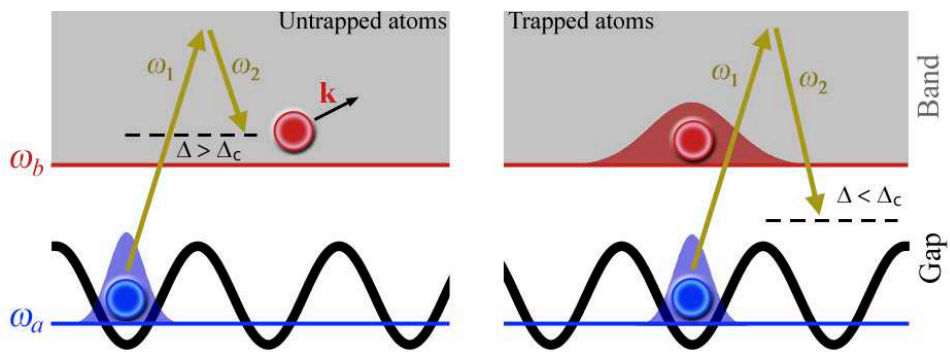}

\caption{Scheme of our proposed setup. Atoms in state $a$ are trapped in an optical lattice, while atoms in state $b$ are free, and can thus have any momentum. An external pair of Raman lasers connect the two levels with some detuning $\Delta$. We will show that above some critical value $\Delta=\Delta_\mathrm{c}$ the atoms in state $b$ are able to leave the trap (left), while below this, they are trapped forming a bound state with atoms in state $a$ (right). This behavior is typical of band--gap systems where the density of states is zero between the connected states.}

\label{Fig1}

\end{figure}

We will show that all this phenomenology can be observed in optical lattices with the setup already presented in \cite{deVega08}, which is depicted in figure \ref{Fig1}. Consider a collection of bosonic atoms with two relevant internal states labeled by $a$ and $b$ (which may correspond to hyperfine ground--state levels, see for example \cite{Jaksch04rev}). Atoms in state $a$ are trapped by a deep optical lattice in which the localized wavefunction of traps at different lattice sites do not overlap (preventing hopping of atoms between sites), while atoms in state $b$ are not affected by the lattice, and hence behave as free particles. A pair of lasers forming a Raman scheme drive the atoms from the trapped state to the free one \cite{Jaksch04rev}, providing an effective interaction between the two types of particles. We consider the situation of having non-interacting bosons in the lattice \cite{Chin10rev}, as well as hard--core bosons in the collisional blockade regime, where only one or zero atoms can be in a given lattice site \cite{Paredes04}. In the first regime, the lattice consists of a collection of harmonic oscillators placed at the nodes of the lattice; in the second regime, two-level systems replace the harmonic oscillators, the two levels corresponding to the absence or presence of an atom in the lattice site. Therefore, it is apparent that this system is equivalent to a collection of independent emitters (harmonic oscillators or atoms) connected only through a common radiation field, the role of this radiation field being played by the free atoms. This system is therefore the cold--atom analog of the quantum--optical systems considered above, with the difference that the radiated particles are massive, and hence have a different dispersion relation than that of photons in vacuum. Moreover, we show that the field of free atoms can be characterized by a dispersion relation which is similar to the one obtained for photons within a photonic band gap material \cite{deVega08,John94,John95}. It is then to be expected, and so we will prove, that this system will show the same kind of phenomenology as its quantum--optical counterparts.

To conclude this introduction let us explain how the article is organized. We first introduce the model and find its associated Hamiltonian (Section \ref{AtomsH}). Then we study the radiative properties of the system when the lattice sites emit independently (Section \ref{SingleDynamics}); we will show in this section that our system shows localization of the free atoms, in analogy to the localization of light occurring in a photonic band gap material, and we identify as well the Markovian and non-Markovian regimes of the emission. In Section \ref{Collective} we study collective effects. We first deduce a reduced master equation for the lattice atoms, explaining under which conditions it is valid. We then show how by changing the system parameters extended Bose--Hubbard and spin models (Section \ref{Extended}), Dicke superradiance of atomic (Section \ref{AtomicSuper}) and harmonic oscillator (Section \ref{HarmonicSuper}) samples, and directional superradiance (Section \ref{Directionality}) can be observed. We finally talk about the effect of restricting the motion of the free atoms to 2D or 1D traps in Section \ref{Dimensionality}, and give some conclusions in Section \ref{Conclusions}.

\section{The model \label{AtomsH}}

Our starting point is the Hamiltonian of the system in second quantization \cite{Greiner96book}. We will denote by $\left\vert a\right\rangle$ and
$\left\vert b\right\rangle$ the trapped and free atomic states, respectively (having internal energies $\hbar\omega_a^0$ and $\hbar\omega_b^0$). Two--body interactions for the trapped atoms are included with the usual contact-like pseudopotential \cite{Bloch08rev, Chin10rev}, but we neglect the collisions for the free atoms. The Hamiltonian is then written as $\hat{H}=\hat{H}_0+\hat{H}_{a-b}$, with
\numparts
\begin{eqnarray}
\fl \hat{H}_0 &= \sum_{j=a,b} \int \mathrm{d}^3\mathbf{r}\, \hat{\Psi}_j^\dagger\left(\mathbf{r}\right) \left(H_j+\hbar\omega_j^0\right)
\hat{\Psi}_j\left(\mathbf{r}\right) + \frac{g}{2} \int \mathrm{d}^3\mathbf{r}\, \hat{\Psi}_a^{\dagger2} \left(\mathbf{r}\right) \hat{\Psi}_a^2\left(\mathbf{r}\right) , \label{H0}
\\
\fl \hat{H}_{a-b} &= \hbar\Omega \int \mathrm{d}^{3}\mathbf{r}\, \mathrm{e}^{\mathrm{i}\left(\mathbf{k}_L\cdot\mathbf{r}-\omega_Lt\right)} \hat{\Psi}_a\left(\mathbf{r}\right) \hat{\Psi}_b^\dagger\left(\mathbf{r}\right) + \mathrm{H.c.}; \label{Hab}
\end{eqnarray}
\endnumparts
$\hat{H}_0$ contains the individual dynamics of the atoms, $H_j$ being the first--quantized motion Hamiltonian of the atom in the corresponding state, and $g=4\pi\hbar^2a_s/m$, where $a_s$ is the s--wave scattering length of the trapped atoms (which have mass $m$). $\hat{H}_{a-b}$ contains the Raman coupling between the atomic states, $\mathbf{k}_L=\mathbf{k}_1-\mathbf{k}_2$ and $\omega_L=\omega_1-\omega_2$ (laser wave vector and frequency in the following) being the relative wave vector and frequency of the two lasers involved in the Raman scheme (see figure \ref{Fig1}), with $\Omega$ the corresponding two--photon Rabi frequency.

For atoms in state $\left\vert a\right\rangle$, $H_a=-\left(\hbar^2/2m\right)\mathbf{\nabla}^2+V_\mathrm{opt}\left(\mathbf{r}\right)$, where $V_\mathrm{opt}\left(\mathbf{r}\right)$ corresponds to a 3--dimensional optical lattice with cubic geometry and lattice period $d_0$. We work with ultracold atoms under conditions such that their wavefunctions can be described by the set of first--band Wannier functions localized around the nodes of the lattice \cite{Jaksch04rev}. The traps of the optical lattice are approximated by isotropic harmonic potentials \cite{Jaksch04rev}, what allows us to write the Wannier functions as
\begin{equation}
w_0\left(\mathbf{r}-\mathbf{r}_\mathbf{j}\right) = \frac{1}{\pi^{3/4}X_0^{3/2}}\exp\left[-\left(\mathbf{r}-\mathbf{r}_\mathbf{j}\right)^{2}/2X_0^2\right],
\end{equation}
where $\mathbf{r}_\mathbf{j}=d_0\mathbf{j}$ is the position of the $\mathbf{j}\in\mathbb{Z}^3$ lattice site (we consider $M$ sites in each orthogonal direction defining the cubic lattice), and $X_0^2=\hbar/m\omega_0$, being $\omega_0$ the frequency of the harmonic trap. The energy associated to
these wave functions is $E_0=3\hbar\omega_0/2$. We will assume that Wannier functions localized at different lattice sites do not overlap, hence
preventing tunneling between sites.

On the other hand, atoms in state $\left\vert b\right\rangle$ can move freely in every direction of space according to $H_b=-\left(\hbar^2/2m\right)\mathbf{\nabla}^2$, and hence the plane waves $\psi_\mathbf{k}\left(\mathbf{r}\right)=\mathrm{e}^{\mathrm{i}\mathbf{k}\cdot\mathbf{r}}/\sqrt{V}$, with energy $E_\mathbf{k}=\hbar^{2}k^{2}/2m$, are their motion eigenfunctions ($V$ is the total available volume for the free atoms, which we might take as infinite for calculations).

We consider two opposite regimes for the interaction between trapped atoms (remember that in (\ref{H0}) any two--body interaction involving the free atoms has been neglected). The first limit consists in neglecting the interactions, which might be accomplished by, e.g., tuning the scattering length with an additional magnetic field through a Feshbach resonance \cite{Chin10rev}. Expanding the quantum fields onto their corresponding motion eigenstates, we get within the interaction picture:
\begin{equation}
\hat{H}_{a-b}=\sum_{\mathbf{j},\mathbf{k}} g_k \mathrm{e}^{\mathrm{i}\Delta_kt-\mathrm{i}\left(\mathbf{k}-\mathbf{k}_{L}\right)\cdot\mathbf{r}_\mathbf{j}} \hat{a}_\mathbf{j}\hat{b}_\mathbf{k}^\dagger+\mathrm{H.c.}, \label{Hoscis}
\end{equation}
with
\begin{equation}
g_k = \hbar\Omega\sqrt{\frac{8\pi^{3/2}X_0^3}{V}}\exp\left[-\frac{1}{2}X_0^2\left(\mathbf{k}-\mathbf{k}_L\right)^{2}\right], \qquad \Delta_k = \frac{\hbar k^2}{2m}-\Delta, \label{Hpar}
\end{equation}
$\Delta=\omega_L-\left(\omega_b-\omega_a\right)$ being the detuning of the laser frequency respect to the $\left\vert a\right\rangle \rightleftharpoons \left\vert b\right\rangle$ transition ($\omega_a=\omega_a^0+3\omega_0/2$ and $\omega_b=\omega_b^0$). The operators $\{\hat{a}_\mathbf{j},\hat{a}_\mathbf{j}^\dagger\}$ and $\{\hat{b}_\mathbf{k},\hat{b}_\mathbf{k}^\dagger\}$ satisfy canonical bosonic commutation relations, and create or annihilate an atom at lattice site $\mathbf{j}$ and a free atom with momentum $\mathbf{k}$, respectively.

As for the second limit, we assume that the on--site repulsive atom--atom interaction is the dominant energy scale, and hence the trapped atoms behave as hard--core bosons in the collisional blockade regime, what prevents the presence of two atoms in the same lattice site \cite{Paredes04}; this means that the spectrum of $\hat{a}_\mathbf{j}^\dagger\hat{a}_\mathbf{j}$ can be restricted to the first two states $\left\{\left\vert 0\right\rangle_\mathbf{j},\left\vert 1\right\rangle_\mathbf{j}\right\}$, having 0 or 1 atoms at site $\mathbf{j}$, and then the boson operators $\left\{\hat{a}_\mathbf{j}^\dagger,\hat{a}_\mathbf{j}\right\}$ can be changed by spin-like ladder operators $\left\{\hat{\sigma}_\mathbf{j}^\dagger,\hat{\sigma}_\mathbf{j}\right\}=\left\{\vert 1\rangle_\mathbf{j}\langle 0\vert, \vert 0\rangle_\mathbf{j}\langle 1\vert\right\}$. In this second limit the Hamiltonian reads
\begin{equation}
\hat{H}_{a-b} = \sum_{\mathbf{j},\mathbf{k}} g_k \mathrm{e}^{\mathrm{i}\Delta_kt-\mathrm{i}\left(\mathbf{k}-\mathbf{k}_L\right)\cdot\mathbf{r}_\mathbf{j}} \hat{\sigma}_\mathbf{j} \hat{b}_\mathbf{k}^\dagger+\mathrm{H.c.}. \label{Hspins}
\end{equation}

Hamiltonians (\ref{Hoscis}) and (\ref{Hspins}) show explicitly how this system mimics the dynamics of collections of harmonic oscillators or atoms,
respectively, interacting with a common radiation field. Note that, particularly, the dispersion relation appearing in (\ref{Hpar}) is similar to that of the radiation field within an anisotropic 3D photonic band--gap material, where photons acquire an effective mass close to the gap \cite{John94,John95}. We will show how by varying the parameters of these Hamiltonians, the system has access to regimes showing many different phenomena, as explained in the Introduction.

Note finally that in order to satisfy that the trapped atoms are within the first Bloch band, it is required that $\omega_0\gg\Delta,\Omega$.

\section{Emission of an atom from a single site \label{SingleDynamics}}

Before studying the collective behavior of the atoms in the lattice, it is convenient to understand the different regimes of emission when the sites
emit independently. To this aim we first study the emission properties of one atom in a single site following the analysis performed in \cite{John94,John95} for an atom embedded in a photonic band--gap material. We are going to show that there exists a critical value of the detuning above which the atom is emitted, while below which the radiated atom gets bound to the trapped atom, and there is a nonzero probability for the atom to remain in the trap (see figure \ref{Fig1}). This second regime is the analog of the photon--atom bound state predicted more than 25 years ago \cite{John84}. We will also identify the Markovian and non-Markovian regimes of the emission.

Following the Weisskopf--Wigner procedure \cite{Weisskopf30,Scully97book}, we write the state of an atom in a single site (which we take as $\mathbf{j}=\mathbf{0}$) as
\begin{equation}
\left\vert\psi\left(t\right)\right\rangle = A\left(t\right) \left\vert1,\left\{0\right\}\right\rangle + \sum_\mathbf{k} B_\mathbf{k}\left(t\right)
\left\vert0,1_\mathbf{k}\right\rangle, \label{State}
\end{equation}
where $\left\vert1,\left\{0\right\}\right\rangle$ refers to the state with an atom in the trap and no free atoms, and $\left\vert0,1_\mathbf{k}\right\rangle$ to the state with no trapped atoms and one free atom with momentum $\mathbf{k}$. We can use the Schr\"{o}dinger equation to write the evolution equations of the coefficients $A$ and $B_\mathbf{k}$; then, the equation of $B_\mathbf{k}$ can be formally integrated arriving to a single equation for $A$ given by
\begin{equation}
\dot{A}\left(t\right) = - \int_0^t \mathrm{d}t^\prime\, G\left(t-t^\prime\right) A\left(t^\prime\right) , \label{NonMarkovianA}
\end{equation}
where we have assumed that the atom is initially in the trap, that is, $A\left(0\right)=1$ and $B_\mathbf{k}\left(0\right)=0$. This is a non-Markovian equation where the free atoms enter the dynamics of the trapped atoms as a reservoir with correlation function
\begin{equation}
G\left(\tau\right) = \frac{1}{\hbar^2} \sum_\mathbf{k} \left\vert g_k\right\vert^2 \mathrm{e}^{-\mathrm{i}\Delta_k\tau} = \frac{\Omega^2}{\left(1+\frac{\mathrm{i}}{2}\omega_0\tau\right)^{3/2}} \exp\left(i\Delta\tau\right) , \label{G}
\end{equation}
where we have taken the continuous limit for the momenta in the last equality and $\mathbf{k}_L=\mathbf{0}$ for simplicity. Note that this correlation function coincides exactly with that of atoms in anisotropic 3D photonic band--gap materials \cite{John94,John95}.

\begin{figure}[t]

\flushright \includegraphics[width=6in]{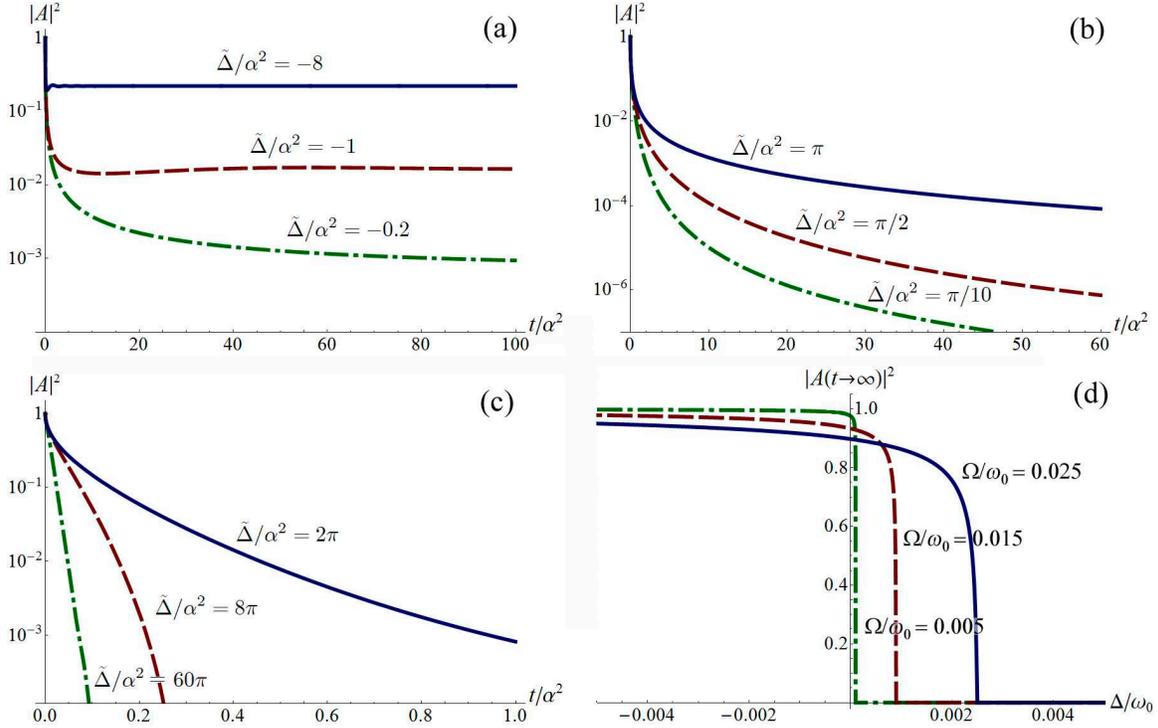}

\caption{(a--c) Time evolution of the population when the lattice sites emit independently. (a) corresponds to the trapping region, (b) to the pure non-Markovian emission of the atoms, and (c) to the radiative region having both Markovian and non-Markovian contributions. The logarithmic scale reveals that the Markovian regime requires $\vert\tilde{\Delta}\vert\gg\pi\alpha^{2}$. (d) Steady state population for finite $\omega_0$ as a function of the detuning $\Delta$ for different values of the Rabi frequency $\Omega$ (see Appendix I). We can appreciate the phase transition from a trapped to a radiative state for $\Delta=4\Omega^2/\omega_0$.}

\label{Fig2}

\end{figure}

In \ref{Appendix I} we show how to handle equation (\ref{NonMarkovianA}) by using Laplace transform techniques. In particular, in \ref{AI evo} we show that in the strong confinement regime $\omega_0\rightarrow\infty$, this equation can be solved as
\begin{equation}
A\left(t\right) = c \mathrm{e}^{\mathrm{i}\left(b^2+\Delta\right)t} + \frac{2\alpha}{\sqrt\pi} \mathrm{e}^{\mathrm{i}\pi/4} \int_0^{+\infty}\mathrm{d}x\, \frac{\sqrt{x}\mathrm{e}^{\left(-x+\mathrm{i}\Delta\right)t}}{\left(-x+\mathrm{i}\tilde{\Delta}\right)^2+\mathrm{i}4\pi\alpha^2x}, \label{Asol}
\end{equation}
with $\alpha^2=8\Omega^4/\omega_0^3$, $\tilde{\Delta}=\Delta-4\Omega^2/\omega_0$ and
\begin{equation}
\left\{
\begin{array}{ll}
c = 2b_+/ \left(b_+-b_-\right) \quad \mathrm{and} \quad b=b_+ & \mathrm{for} \quad \tilde{\Delta}<0,
\\
c = 0 & \mathrm{for} \quad 0<\tilde{\Delta}<\pi\alpha^2,
\\
c=2b_-/\left(b_--b_+\right) \quad \mathrm{and} \quad b=b_- & \mathrm{for} \quad \tilde{\Delta}>\pi\alpha^2,
\end{array}
\right.
\end{equation}
where $b_\pm=\sqrt{\pi}\alpha\left(-1\pm\sqrt{1-\tilde{\Delta}/\pi\alpha^2}\right)$.

Solution (\ref{Asol}) has a very suggestive form. The first term has exponential evolution as expected for a Markovian radiation process, while the
integral term has a nontrivial time dependence which always goes to zero for sufficiently large times. The time while the integral term is comparable to the exponential one defines the non-Markovian part of the evolution. This result also shows that the behavior of the emission is highly dependent on the parameters of the system:

\begin{itemize}
\item In the $\tilde{\Delta}<0$ region $\mathrm{Im}\left\{b_+^2\right\}=0$, and hence the steady state population is nonzero and given by $\vert A(t\rightarrow\infty)\vert^2=\vert c\vert^2=\left(1-1/\sqrt{1+\vert\tilde{\Delta}\vert/\pi\alpha^2}\right)^2$; a fraction $1-\vert c\vert^2$ is radiated during the non-Markovian period which lasts longer as $\tilde{\Delta}/\alpha^{2}$ goes to zero (see figure \ref{Fig2}a). In this region the radiated atoms are emitted in the form of evanescent waves, and the equivalent of a photon--atom bound state \cite{John84} is formed. \vspace{6pt}

\item In the $0<\tilde{\Delta}<\pi\alpha^2$ region (figure \ref{Fig2}b) the steady state population is zero, i.e., the atom leaves the trap eventually, and the evolution is dictated solely by the integral part of the solution (hence it is pure non-Markovian radiation). \vspace{6pt}

\item Finally, in the $\tilde{\Delta}>\pi\alpha^2$ region the atom is radiated in a Markovian fashion only for $\tilde{\Delta}\gg\alpha^2$, with a decay rate given by $\Gamma_0=\mathrm{Im}\left\{b_-^2\right\}\approx4\Omega^2\sqrt{2\pi\tilde{\Delta}/\omega_0^3}$; if this is not the case ($\tilde{\Delta}\sim\alpha^{2}$), the integral part is comparable to the exponential part during most of the evolution time, and hence the radiation process is non-Markovian (figure \ref{Fig2}c).
\end{itemize}

Therefore, there exists a phase transition at $\Delta=4\Omega^2/\omega_0$ in close analogy to that of spontaneous emission in photonic
band--gap materials \cite{John94,John95}: Above this value the atom is forced to leave the trap, while below it there is a nonzero trapped population left in the steady state. Using the techniques we explain in \ref{AI ss}, it is possible to show that this phase transition is also present for finite $\omega_0$ (see figure \ref{Fig2}d).

Note that the Markovianity condition $\vert\tilde{\Delta}\vert\gg\alpha^2$ can be recasted as $\vert\tilde{\Delta}\vert\gg\Gamma_0$, which coincides with the usual Markov limit $\tau_\mathrm{S}\gg\tau_\mathrm{R}$ \cite{Breuer02book} once the characteristic evolution times of the lattice atoms and the reservoir's correlation function are identified with $\tau_\mathrm{S}=\Gamma_0^{-1}$ and $\tau_\mathrm{R}=\vert\tilde{\Delta}\vert^{-1}$, respectively.

\section{Collective dynamics \label{Collective}}

Let us study now the collective emission of atoms from many sites. We analyze this through the master equation for the reduced density operator of the lattice  within the Born--Markov approximation \cite{Breuer02book}.

We introduce the Born approximation by fixing the state of the reservoir to a non-evolving vacuum. This leads to an effective interaction between different lattice sites mediated by the two--point correlation function
\begin{equation}
\fl G_{\mathbf{j}-\mathbf{l}}\left(\tau\right) = \frac{1}{\hbar^2} \sum_{\mathbf{k}} g_k^2 \mathrm{e}^{-\mathrm{i}\Delta_k\tau+\mathrm{i}\left(\mathbf{k}-\mathbf{k}_{L}\right)\cdot\mathbf{r}_{\mathbf{j}-\mathbf{l}}}
\left\langle\hat{b}_\mathbf{k}\hat{b}_\mathbf{k}^\dagger\right\rangle = \exp\left(\frac{-r_{\mathbf{j}-\mathbf{l}}^2/4X_0^2}{1+\mathrm{i}\omega_{0}\tau/2}-\mathrm{i}\mathbf{k}_{L}\cdot\mathbf{r}_{\mathbf{j}-\mathbf{l}}\right)
G\left(\tau\right), \label{Gjl}%
\end{equation}
being $\mathbf{r}_{\mathbf{j}-\mathbf{l}}=\mathbf{r}_\mathbf{j}-\mathbf{r}_\mathbf{l}=d_{0}\left(\mathbf{j}-\mathbf{l}\right)$. The Born approximation yields the dynamics of the system to the leading order in the coupling parameters $g_k$ \cite{Breuer02book}. On the other hand, as commented at the end of the previous section, the Markov approximation is valid only if $\vert\tilde{\Delta}\vert\gg\Gamma$, where $\Gamma$ is a typical evolution rate for trapped atoms, which can differ from $\Gamma_0$ owed to a renormalization due to collective effects as we show below. We also include the single--site energy shift that we found in the previous section (the analogous of the Lamb shift in the optical case), and substitute $\Delta$ by $\tilde{\Delta}$ in the two--point correlation function.

\begin{figure}[t]

\flushright \includegraphics[width=6in]{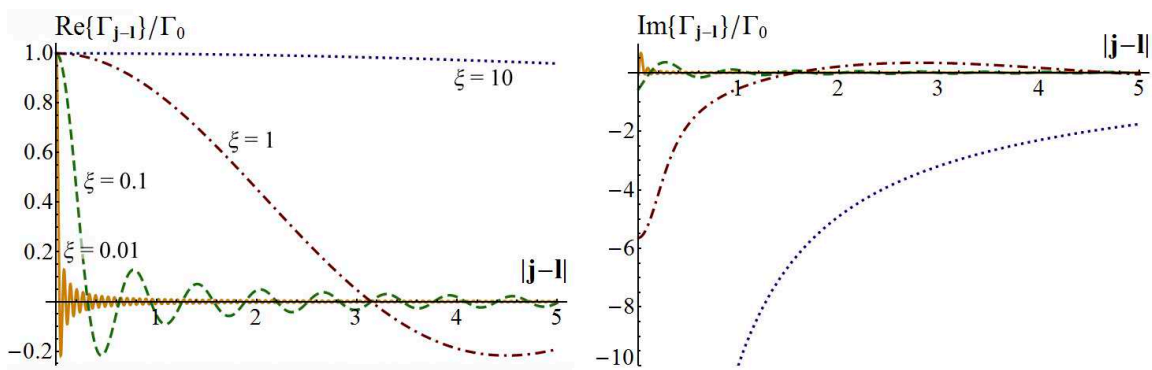}

\caption{We plot the Markov couplings (real and imaginary parts at left and right, respectively) as a function of the distance between sites for $\tilde{\Delta}>0$. We have chosen $d_0/X_0=10$ and $\mathbf{k}_L=\mathbf{0}$, and plotted 4 different values of $\xi$. It can be appreciated that this parameter controls the spatial range of the interactions.}

\label{Fig3}

\end{figure}

The master equation for the reduced density operator of the trapped atoms $\hat{\rho}$ is found by using standard techniques from the theory of open quantum systems \cite{Breuer02book}, and reads
\begin{equation}
\frac{\mathrm{d}\hat{\rho}}{\mathrm{d}t} = \sum_{\mathbf{j},\mathbf{l}} \Gamma_{\mathbf{j}-\mathbf{l}} \hat{a}_\mathbf{l} \hat{\rho} \hat{a}_\mathbf{j}^\dagger - \Gamma_{\mathbf{j}-\mathbf{l}} \hat{a}_\mathbf{j}^\dagger \hat{a}_\mathbf{l} \hat{\rho} + \mathrm{H.c.}; \label{Master}
\end{equation}
a similar equation is obtained for the hard--core bosons but replacing the boson operators by the corresponding spin operators. The effective interaction between different lattice sites is mediated by the Markov couplings (see \ref{Appendix II})
\begin{eqnarray}
\fl \Gamma_{\mathbf{j}-\mathbf{l}} = \mathrm{i} \exp\left(-\mathrm{i}\mathbf{k}_{L}\cdot\mathbf{r}_{\mathbf{j}-\mathbf{l}}\right) \frac{\Gamma_0\xi}{\left\vert\mathbf{j}-\mathbf{l}\right\vert} \left[ 1 - \mathrm{erf}\left(\frac{d_0}{2X_0}\left\vert\mathbf{j}-\mathbf{l}\right\vert\right) - \exp\left(-\nu\frac{\left\vert\mathbf{j}-\mathbf{l}\right\vert}{\xi}\right) \right], \label{Couplings}
\end{eqnarray}
where $\xi=1/d_0k_0$ with $k_0=X_0^{-1}\sqrt{2\vert\tilde{\Delta}\vert/\omega_0}$, and $\nu=1$ $(-\mathrm{i})$ for $\tilde{\Delta}<0$ $(\tilde{\Delta}>0)$. The error function is defined as $\mathrm{erf}\left(x\right)=\left(2/\sqrt{\pi}\right)\int_0^x\mathrm{d}u\exp\left(-u^2\right)$. This expression has been evaluated in the strong confinement regime $\omega_0\rightarrow\infty$ and considering $k_L\ll X_0^{-1}$ (see \ref{Appendix II}). Note that the term `$1-\mathrm{erf}\left(d_0\left\vert\mathbf{j}-\mathbf{l}\right\vert/2X_0\right)$' is basically zero for $\mathbf{j}\neq\mathbf{l}$, and therefore the $\xi$ parameter dictates the spatial range of the interactions as can be appreciated in figure \ref{Fig3}. Finally, we remind that $\Gamma_0=4\Omega^2\sqrt{2\pi\vert\tilde{\Delta}\vert/\omega_0^3}$ is the single--emitter decay rate.

\subsection{Extended Bose--Hubbard and spin models \label{Extended}}

For $\mathbf{k}_L=0$ and $\tilde{\Delta}<0$ the Markov couplings are purely imaginary and have negative imaginary parts. Under this conditions, the master equation takes a Hamiltonian form with effective Hamiltonians
\begin{equation}
\hat{H}_\mathrm{eff} = -\sum_{\mathbf{j},\mathbf{l}} \hbar \left\vert\Gamma_{\mathbf{j}-\mathbf{l}}\right\vert \hat{a}_\mathbf{j}^\dagger \hat{a}_\mathbf{l} \quad \mathrm{and} \quad \hat{H}_\mathrm{eff} = -\sum_{\mathbf{j},\mathbf{l}} \hbar \left\vert\Gamma_{\mathbf{j}-\mathbf{l}}\right\vert \hat{\sigma}_\mathbf{j}^\dagger \hat{\sigma}_\mathbf{l},
\end{equation}
for non-interacting and hard--core bosons, respectively. The first Hamiltonian describes extended hopping in the lattice, a feature that has been recently shown to be helpful for achieving true incompressible Mott phases which otherwise can be blurred because of the additional slowly varying harmonic trap used to confine the atoms within the lattice \cite{Rousseau10}. The second Hamiltonian is equivalent to an extended ferromagnetic Ising-like Hamiltonian, and connects our system to extended spin models. Note that for $\mathbf{j}\neq\mathbf{l}$ the couplings have a Yukawa form
\begin{equation}
\left\vert\Gamma_{\mathbf{j}-\mathbf{l}}\right\vert \approx \frac{\Gamma_0}{\left\vert\mathbf{j}-\mathbf{l}\right\vert/\xi} \exp\left(-\frac{\left\vert\mathbf{j}-\mathbf{l}\right\vert}{\xi}\right) ,
\end{equation}
and can even take a Coulomb form if $\xi$ is large enough. These kind of interactions (specially the Coulomb-like) are difficult to obtain with other techniques.

These results show that in the trapping region the system under study could be useful for quantum simulation of condensed--matter
phenomena requiring extended interactions.

\subsection{Dicke superradiance \label{Superradiance}}

In this section we connect our system to superradiant Dicke-like phenomena both in atomic and harmonic oscillator samples (we will still take
$\mathbf{k}_L=0$).

For $\tilde{\Delta}>0$, the Markov couplings are complex in general, and therefore the master equation of the system takes the form
\begin{equation}
\frac{\mathrm{d}\hat{\rho}}{\mathrm{d}t} = \frac{1}{\mathrm{i}\hbar} \left[\hat{H}_\mathrm{d},\hat{\rho}\right] + \mathcal{D}\left[\hat{\rho}\right] , \label{Master Eq}
\end{equation}
with a dissipation term given by
\begin{equation}
\mathcal{D}\left[\hat{\rho}\right] = \sum_{\mathbf{j},\mathbf{l}} \gamma_{\mathbf{j}-\mathbf{l}} \left( 2\hat{a}_\mathbf{l}\hat{\rho}\hat{a}_\mathbf{j}^\dagger - \hat{a}_\mathbf{j}^\dagger\hat{a}_\mathbf{l}\hat{\rho} - \hat{\rho}\hat{a}_\mathbf{l}^\dagger\hat{a}_\mathbf{j} \right) , \label{Diss}
\end{equation}
having collective decay rates
\begin{equation}
\gamma_{\mathbf{j}-\mathbf{l}} = \mathrm{Re}\left\{\Gamma_{\mathbf{j}-\mathbf{l}}\right\} = \Gamma_0 \mathrm{sinc}\left(\frac{\left\vert\mathbf{j}-\mathbf{l}\right\vert}{\xi}\right) ,
\end{equation}
and a reversible term corresponding to inhomogeneous dephasing with Hamiltonian
\begin{equation}
\hat{H}_\mathrm{d} = \sum_{\mathbf{j},\mathbf{l}} \hbar \Lambda_{\mathbf{j}-\mathbf{l}} \hat{a}_\mathbf{j}^\dagger \hat{a}_\mathbf{l}, \label{Hd}
\end{equation}
being
\begin{equation}
\Lambda_{\mathbf{j}-\mathbf{l}} = \mathrm{Im}\left\{\Gamma_{\mathbf{j}-\mathbf{l}}\right\} = \frac{\Gamma_0\xi}{\left\vert\mathbf{j}-\mathbf{l}\right\vert} \left[ 1 - \mathrm{erf}\left(\frac{d_0}{2X_0}\left\vert\mathbf{j}-\mathbf{l}\right\vert\right) - \cos\left(\frac{\left\vert\mathbf{j}-\mathbf{l}\right\vert}{\xi}\right) \right]. \label{Lambda}
\end{equation}
The same holds for hard--core bosons but replacing the boson operators by the corresponding spin operators.

For $\xi\ll1$ the Markov couplings do not connect different lattice sites, that is $\Gamma_{\mathbf{j}-\mathbf{l}}\simeq\Gamma_\mathbf{0}\delta_{\mathbf{j},\mathbf{l}}$, and the sites emit independently. On the other hand, when $\xi\gg M$ the collective decay rates become homogeneous, $\gamma_{\mathbf{j}-\mathbf{l}}\simeq\Gamma_0$, and we enter the Dicke regime. Hence, we expect to observe the superradiant phase--transition in our system by varying the parameter $\xi$.

Note that in the Dicke regime the dephasing term cannot be neglected and connects the sites inhomogeneously with $\Lambda_{\mathbf{j}\neq\mathbf{l}}\simeq\Gamma_0\xi/\left\vert\mathbf{j}-\mathbf{l}\right\vert$. This term appears in the optical case too, although it was inappropriately neglected in the original work by Dicke \cite{Dicke54} when assuming the dipolar approximation in his initial Hamiltonian, and slightly changes his original predictions as pointed out in \cite{Friedberg72,Friedberg74} (see also \cite{Gross82rev}).

In the following we analyze the superradiant behavior of our system by studying the evolution of the total number of particles in the lattice
$n_\mathrm{T}=\sum_\mathbf{j}\left\langle\hat{a}_\mathbf{j}^\dagger\hat{a}_\mathbf{j}\right\rangle$ and the rate of emitted atoms
\begin{equation}
\mathcal{R}\left(t\right) = \sum_\mathbf{k} \frac{\mathrm{d}}{\mathrm{d}t} \left\langle\hat{b}_\mathbf{k}^\dagger \hat{b}_\mathbf{k}\right\rangle = -\frac{\mathrm{d}n_\mathrm{T}}{\mathrm{d}t};
\end{equation}
in the last equality we have used that the total number operator $\sum_\mathbf{k}\hat{b}_\mathbf{k}^\dagger\hat{b}_\mathbf{k}+\sum_\mathbf{j}\hat{a}_\mathbf{j}^\dagger\hat{a}_\mathbf{j}$ is a constant of motion.

\subsubsection{Hard--core bosons: Atomic superradiance. \label{AtomicSuper}}

Let us start by analyzing the case of a lattice in an initial Mott phase having one atom per site in the collisional blockade regime, which is the analog of an ensemble of excited atoms \cite{deVega08}. As explained in the Introduction, Dicke predicted that superradiance should appear in this system as an enhancement of the emission rate at early times \cite{Dicke54}, although this was later proved to happen only if the effective number of interacting emitters exceeds some threshold value \cite{Ernst68,Agarwal70,Rehler71}: This is the superradiant phase transition.

\begin{figure}[t]

\flushright \includegraphics[width=6in]{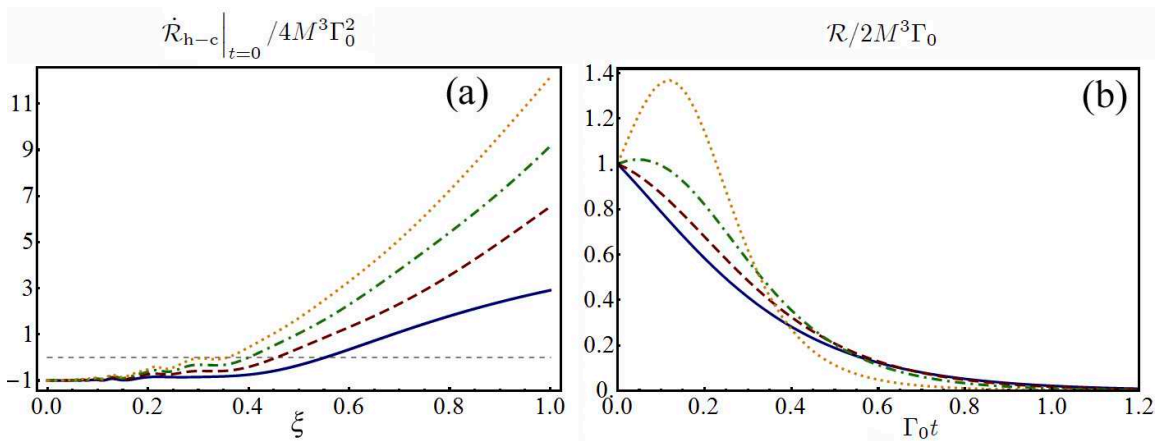}

\caption{Collective emission properties for an initial Mott state of hard--core bosons. (a) Time derivative of the rate at $t=0$ as a function of the range of the interactions $\xi$, see (\ref{HardCoreRateDerIni}). The values $M^3=8$ (solid, blue), $27$ (dashed, red), $64$ (dashed--dotted, green), and $125$ (dotted, yellow) are considered. It can be appreciated that there exists a critical value of $\xi$ above which the rate is enhanced at the initial times. (b) Rate as a function of time for $M^3=27$. The values $\xi=0.01$ (solid, blue), $0.5$ (dashed, red), $1$ (dashed--dotted, green), and $10$ (dotted, yellow) are considered. As expected from (a), the maximum of the rate is delayed above some critical $\xi$ value. Note that both the rate and its derivative have been normalized to the values expected for independent emitters, which are $4M^3\Gamma_0^2$ and $2M^3\Gamma_0$, respectively.}

\label{Fig4}

\end{figure}

In our system, the number of interacting spins is governed by the parameter $\xi$ (see figure \ref{Fig3}), and the simplest way to show that the superradiant phase transition appears by varying it, is by evaluating the initial slope of the rate which can be written as
\footnote{
Note that the evolution equation of the expectation value of any operator $\hat{O}$ can be written as
\begin{equation}
\frac{\mathrm{d}}{\mathrm{d}t}\left\langle\hat{O}\left(t\right)\right\rangle = \mathrm{tr}\left\{\frac{\mathrm{d}\hat{\rho}}{\mathrm{d}t}\hat{O}\right\} = - \sum_{\mathbf{m},\mathbf{l}} \left\{\Gamma_{\mathbf{m}-\mathbf{l}} \left\langle\left[\hat{O},\hat{a}_\mathbf{m}^\dagger\right] \hat{a}_\mathbf{l}\right\rangle+\Gamma_{\mathbf{m}-\mathbf{l}}^* \left\langle \hat{a}_\mathbf{l}^\dagger\left[\hat{a}_\mathbf{m},\hat{O}\right]\right\rangle\right\}, \label{ExpEvo}
\end{equation}
and similarly the hard--core bosons in terms of the spin ladder operators.
}
\begin{equation}
\left.\frac{\mathrm{d}}{\mathrm{d}t}\mathcal{R}\right\vert_{t=0} = -4M^3 \Gamma_0^2 \left[1-\sum_{\mathbf{m}\neq\mathbf{j}}\frac{\mathrm{sinc}^2\left(\left\vert\mathbf{j}-\mathbf{m}\right\vert/\xi\right)}{M^3}\right] . \label{HardCoreRateDerIni}
\end{equation}

This expression has a very suggestive form: The term corresponding to the rate associated to independent emitters is balanced by a collective contribution arising from the interactions between them. In figure \ref{Fig4}a we show the dependence of this derivative with $\xi$ for various values of the number of sites $M^3$. It can be appreciated that there exists a critical value of $\xi$ above which the sign of the derivative is reversed; hence, the rate increases at the initial time and we expect its maximum to be no longer at $t=0$, which is a signature of superradiance.

The time evolution of the rate for a cubic lattice with $M^3=27$ sites is shown in figure \ref{Fig4}b for different values of $\xi$. We can appreciate how above some critical $\xi$ value the maximum rate of emission is delayed as expected. In order to find $\mathcal{R}(t)$ we have simulated the evolution equations for the coherences $c_\mathbf{jl}=\langle\hat{\sigma}_\mathbf{j}^\dagger\hat{\sigma}_\mathbf{l}\rangle$ and the populations $s_\mathbf{j}=\left\langle\hat{\sigma}_\mathbf{j}^3\right\rangle$, which we close by using the semiclassical approximation $\left\langle\hat{\sigma}_\mathbf{m}^\dagger\hat{\sigma}_\mathbf{j}^3\hat{\sigma}_\mathbf{l}\right\rangle = \left\langle\hat{\sigma}_\mathbf{m}^\dagger\hat{\sigma}_\mathbf{l}\right\rangle\left\langle\hat{\sigma}_\mathbf{j}^3\right\rangle - 2\delta_\mathbf{jl}\left\langle\hat{\sigma}_\mathbf{m}^\dagger\hat{\sigma}_\mathbf{l}\right\rangle$; they read (\ref{ExpEvo})
\numparts
\begin{eqnarray}
\dot{c}_\mathbf{jl} &= -4 \Gamma_0 c_\mathbf{jl} + \sum_\mathbf{m} \Gamma_{\mathbf{l}-\mathbf{m}} c_\mathbf{jm} s_\mathbf{l} + \Gamma_{\mathbf{j}-\mathbf{m}}^\ast c_\mathbf{ml} s_\mathbf{j} ,
\\
\dot{s}_\mathbf{j} &= -2 \sum_\mathbf{l} \Gamma_{\mathbf{j}-\mathbf{l}} c_\mathbf{jl} + \Gamma_{\mathbf{j}-\mathbf{l}}^\ast c_\mathbf{lj}.
\label{HardCoreSemiclassEqs}
\end{eqnarray}
\endnumparts
We have checked the validity of these semiclassical equations by comparing them with a direct simulation of the master equation for small number of sites in 1D and 2D geometries; except for small quantitative deviations, they offer the same results.

\subsubsection{Non-interacting bosons: Harmonic oscillators superradiance. \label{HarmonicSuper}}

Let us analyze now the case of having non-interacting bosons in the lattice, which is equivalent to a collection of harmonic oscillators as discussed before. In previous works on superradiance this system was studied in parallel to its atomic counterpart \cite{Agarwal70}, and here we show how our system offers a physical realization of it. We will show that superradiant effects can be observed in the evolution of the total number of atoms in the lattice, both for initial Mott and superfluid phases
\footnote{
Let us note that a superfluid state with $N$ excitations distributed over the entire lattice is more easily defined in the discrete Fourier--transform (DFT) basis
\begin{equation}
\hat{f}_\mathbf{q}=\frac{1}{M^{3/2}}\sum_\mathbf{j}\exp\left(\frac{2\pi\mathrm{i}}{M}\mathbf{q}\cdot\mathbf{j}\right) \hat{a}_\mathbf{j}, \label{DFT}
\end{equation}
with $\mathbf{q}=\left(q_x,q_y,q_z\right)$ and $q_{x,y,z}=0,1,2,...M-1$, as the state having $N$ excitations in the zero--momentum mode, that is, $\left\vert\mathrm{SF}\right\rangle _N=\left(N!\right)^{-1/2}\hat{f}_\mathbf{0}^{\dagger N}\left\vert0\right\rangle$.
}.

\begin{figure}[t]

\flushright \includegraphics[width=6in]{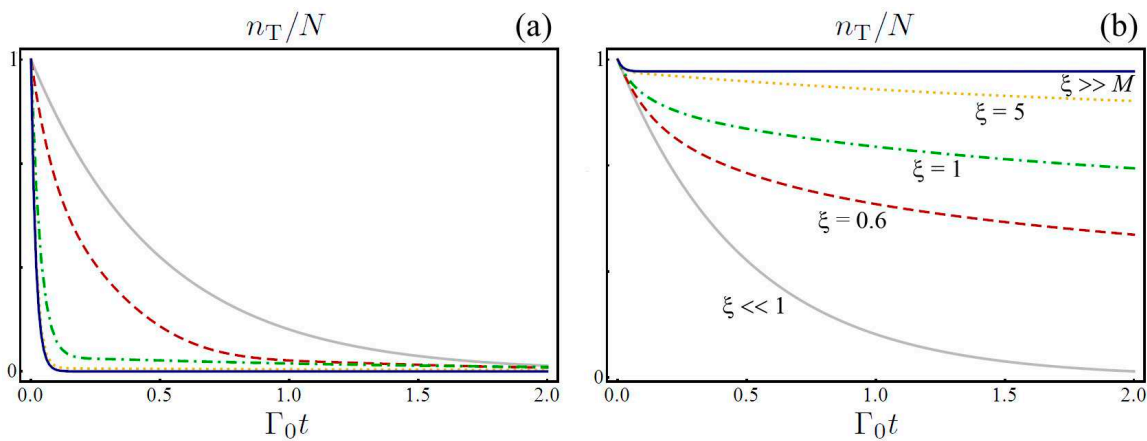}

\caption{Evolution of the total number of atoms in a lattice having $M^3=27$ sites for initial superfluid (a) and Mott (b) phases with $N$ initial
non-interacting atoms. The solid curves correspond to the limits $\xi\ll1$ (grey) and $\xi\gg M$ (dark--blue). Note how the `evaporation time' is reduced for the initial superfluid state as $\xi$ increases (a). Equivalently, note how for an initial Mott state the atoms tend to stay in the lattice as $\xi$ increases (b).}

\label{Fig5}

\end{figure}

The evolution of the total number of atoms in the lattice is given by --see (\ref{ExpEvo})--
\begin{equation}
\dot{n}_\mathrm{T}=-2\sum_{\mathbf{j},\mathbf{l}} \gamma_{\mathbf{j}-\mathbf{l}} \mathrm{Re}\left\{\left\langle\hat{a}_\mathbf{j}^\dagger\hat{a}_\mathbf{l}\right\rangle\right\},
\end{equation}
and hence depends only on the real part of the Markov couplings. Therefore, we restrict our analysis to the dissipative term $\mathcal{D}\left[\hat{\rho}\right]$ of the master equation (\ref{Master Eq}).

By diagonalizing the real, symmetric collective decay rates with an orthogonal matrix $S$ such that $\sum_\mathbf{jl}S_\mathbf{pj}\gamma_{\mathbf{j}-\mathbf{l}}S_\mathbf{ql}=\bar{\gamma}_\mathbf{p}\delta_\mathbf{pq}$, one can find a set of modes $\left\{\hat{c}_\mathbf{p}=\sum_\mathbf{j}S_\mathbf{pj}\hat{a}_\mathbf{j}\right\}$ with definite decay properties. Then, it is completely straightforward to show that the total number of atoms can be written as a function of time as
\begin{equation}
n_\mathrm{T}\left(t\right) = \sum_\mathbf{p} \left\langle\hat{c}_\mathbf{p}^\dagger\hat{c}_\mathbf{p}\left(0\right)\right\rangle \exp\left(-2\bar{\gamma}_\mathbf{p}t\right). \label{ntt}
\end{equation}

In general, $\gamma_{\mathbf{j}-\mathbf{l}}$ requires numerical diagonalization. However, in the limiting cases $\xi\ll1$ and $\xi\gg M$, its spectrum becomes quite simple. Following the discussion after (\ref{Lambda}), in the $\xi\ll1$ limit $\gamma_{\mathbf{j}-\mathbf{l}}=\Gamma_0\delta_\mathbf{jl}$ is already diagonal and proportional to the identity. Hence, any orthogonal matrix $S$ defines an equally suited set of modes all decaying with rate $\Gamma_0$. Therefore, if the initial number of atoms in the lattice is $N$, this will evolve as
\begin{equation}
n_\mathrm{T}\left(t\right) = N \exp\left(-2\Gamma_0t\right), \label{TotalNumberIndependent}
\end{equation}
irrespective of the particular initial state of the lattice (e.g., Mott or superfluid). The emission rate $\mathcal{R}=2\Gamma_0N\exp\left(-2\Gamma_0t\right)$, corresponds to the independent decay of the $N$ atoms as expected in this regime having no interaction between the emitters.

Let us consider now the opposite limit $\xi\gg M$; in this case $\gamma_{\mathbf{j}-\mathbf{l}}=\Gamma_0$ $\forall$ $\left(\mathbf{j},\mathbf{l}\right)$, and the dissipative term can be written in terms of the symmetrical DFT mode only as $\mathcal{D}\left[\hat{\rho}\right] = M^3\Gamma_0\left(2\hat{f}_\mathbf{0}\hat{\rho}\hat{f}_\mathbf{0}^\dagger - \hat{f}_\mathbf{0}^\dagger\hat{f}_\mathbf{0}\hat{\rho} - \hat{f}_\mathbf{0}^\dagger\hat{f}_\mathbf{0}\hat{\rho}\right)$, see (\ref{DFT}). Hence, the DFT basis diagonalizes the problem, and shows that all the modes have zero decay rate except the symmetrical one, which has an enhanced rate proportional to the number of emitters. Therefore, starting with a superfluid state, the $N$ initial atoms will decay exponentially with initial rate $NM^3\Gamma_0$, that is
\begin{equation}
n_\mathrm{T}\left(t\right) = N\exp\left(-2M^3\Gamma_0t\right). \label{TotalNumberSuperfluid}%
\end{equation}
On the other hand, if the initial state corresponds to a Mott phase, most of the atoms will remain in the lattice, as only the component which projects onto the symmetric mode will be emitted; concretely, from (\ref{ntt}) and (\ref{DFT}) the number of atoms in the lattice will evolve for this particular initial state as
\begin{equation}
n_\mathrm{T}\left(t\right) = \left(N-\frac{N}{M^3}\right) + \frac{N}{M^3} \exp\left(-2M^3\Gamma_0t\right). \label{TotalNumberMott}
\end{equation}

Hence, according to this picture, superradiant collective effects can be observed in our system by two different means. Calling $t_{0}$ the time needed to radiate the atoms in the absence of collective effects, one could start with a superfluid phase and measure this `evaporation time' as a function of $\xi$; this should go from $t_0$ for $\xi\ll1$, to a much shorter time $t_0/M^3$ for $\xi\gg M$ (see figure \ref{Fig5}a). Alternatively, one could start with a Mott phase, and measure the number of atoms left in the lattice in the steady state as a function of $\xi$; in this case, it should go from $n_\mathrm{T,steady}=0$ after a time $t_0$ for $\xi\ll1$, to $n_\mathrm{T,steady}=N-N/M^3$ after a time $t_0/M^3$ for $\xi\gg M$ (see figure \ref{Fig5}b).

In order to find the evolution of $n_\mathrm{T}$ we have simulated the equations satisfied by the coherences $c_\mathbf{jl}=\langle\hat{a}_\mathbf{j}^\dagger\hat{a}_\mathbf{l}\rangle$, which read --see (\ref{ExpEvo})--
\begin{equation}
\dot{c}_\mathbf{jl} = -\sum_\mathbf{m} \left[ \Gamma_{\mathbf{l}-\mathbf{m}} c_\mathbf{jm} + \Gamma_{\mathbf{j}-\mathbf{m}}^\ast c_\mathbf{ml}\right]. \label{cSystem}
\end{equation}
Note that in this case the equations are closed without the need of a semiclassical approximation, and hence they are exact. Note that they reproduce the analytic evolution of $n_\mathrm{T}$ as given by (\ref{TotalNumberIndependent}), (\ref{TotalNumberMott}), and (\ref{TotalNumberSuperfluid}) in the corresponding limits (see figure \ref{Fig5}).

Our results connect directly to those found by Agarwal some decades ago \cite{Agarwal70}. Working with a Dicke-like model, he showed that if all the
oscillators start in the same coherent state $\left\vert\alpha\right\rangle $, the initial number of excitations, which in that case is given by
$N=M^3\left\vert\alpha\right\vert^2$, decays following (\ref{TotalNumberSuperfluid}). This is not a coincidence, but rather a consequence that, if $N$ is large enough, a multi--coherent state of that kind is a good approximation of a superfluid state with that number of excitations. He also predicted that if the oscillators start in a number state, most of the excitations would remain in the steady state as follows from (\ref{TotalNumberMott}).

\subsection{Directional superradiance \label{Directionality}}

The last phenomenon connected to quantum--optics that we want to discuss is how allowing $\mathbf{k}_L\neq0$ the direction of emission of atoms can be controlled.

This follows from the fact that by tuning the laser wave vector such that $k_L=k_0$ while working in the $\tilde{\Delta}>0$ regime, the master equation of our system (\ref{Master}) has exactly the same form as that of \cite{Porras07} (see also \cite{Porras07a}), where the properties of light emitted by regular arrays of atoms were analyzed in terms of the rate and the direction of emission. Hence, we could expect the same phenomena to appear in our system, with the difference that in our case the emitted particles are atoms and not photons. Let us explain the results found in \cite{Porras07} and translate them to our system.

We are particularly interested in the analysis performed in \cite{Porras07} of the emission properties of an initial symmetric Dicke state (a symmetrical spin wave in its notation). It was shown that the average number of photons emitted in the direction of the vector $\mathbf{u}_\Omega=\left(\sin\theta\cos\phi,\sin\theta\sin\phi,\cos\theta\right)$, $\theta$ and $\phi$ being the polar and azimuthal spherical angles, respectively, can be written as
\begin{equation}
I\left(\Omega\right) = \frac{1}{4\pi M^3} \frac{\Gamma_0}{\Gamma} \sum_{\mathbf{j},\mathbf{l}} \exp\left[\mathrm{i}\left(k_L\mathbf{u}_\Omega-\mathbf{k}_L\right)\cdot\mathbf{r}_{\mathbf{j}-\mathbf{l}}\right] \label{Iomega}
\end{equation}
where the total emission rate $\Gamma$ is found from the normalization condition $\int\mathrm{d}\Omega I\left(\Omega\right)=1$ ($\Omega$ is the solid angle).

This expression was deduced in \cite{Porras07} in the $M\gg1$ limit by diagonalizing the master equation with the use of periodic boundary conditions. The limit $\Gamma\ll c/L$ was considered ($c$ is the speed of light and $L=Md_0$ is the length of the lattice in a given direction), that is, it was assumed that the emission time is larger than the time taken by a photon to leave the sample. Following the same steps, but taking into account the different dispersion relation of the emitted particles in our system, the same expression (\ref{Iomega}) can be proved but now under the condition $\Gamma\ll v_0/L$, where $v_0=\hbar k_0/m$ is the `resonant' speed of the emitted atoms. We will explain latter what this condition means in terms of the system parameters.

In the optical case treated in \cite{Porras07}, the angular distribution (\ref{Iomega}) applies strictly only to the single excitation sector. In our case, this result is specially relevant for the case of non-interacting atoms trapped in the optical lattice in an initial superfluid state: since a superfluid state can be described as a state of $N$ atoms in the completely symmetric state, see (\ref{DFT}), the results of \cite{Porras07} imply here that those $N$ trapped atoms will emit $N$ free atoms with an angular distribution given by (\ref{Iomega}).

The sums in (\ref{Iomega}) can be performed analytically, leading to the result
\begin{equation}
I\left(\Omega\right) = \frac{1}{4\pi M^3} \frac{\Gamma_0}{\Gamma} \prod_{\alpha=x,y,z} \frac{\sin^2\left[\left(\mathbf{u}_\Omega-\mathbf{\hat{k}}_L\right)_\alpha M/2\xi\right]} {\sin^2\left[\left(\mathbf{u}_\Omega-\mathbf{\hat{k}}_L\right)_\alpha/2\xi\right]},
\end{equation}
where we have denoted by $\mathbf{\hat{k}}_L=\mathbf{k}_L/k_L$ the unit vector in the `laser' direction. This function has a diffraction maximum whenever the denominator vanishes, that is, for vectors $\mathbf{m}\in\mathbb{Z}^3$ such that $\mathbf{u}_\Omega=\mathbf{\hat{k}}_L+2\pi\xi\mathbf{m}$. In order to analyze directionality it is better to rewrite this condition as $\pi\xi\vert\mathbf{m}\vert=\vert\mathbf{\hat{k}}_{L}\cdot\mathbf{\hat{m}}\vert$, where $\mathbf{\hat{m}}=\mathbf{m}/\vert\mathbf{m}\vert$. Then we see that for $\pi\xi>1$ only $\mathbf{m}=\mathbf{0}$ can satisfy this condition, and hence the atoms are mostly emitted in the laser direction. This is the directional regime.

The emission rate $\Gamma$ and the angular width of the atomic beam $\Delta\theta$ must be evaluated numerically in general. However,
approximating the diffraction peaks by Gaussian functions, a limit which is justified if the peaks are narrow enough ($\xi/M\ll1$), we can get the
following expression for them \cite{Porras07}
\begin{equation}
\Gamma=\chi\Gamma_0 \quad \mathrm{with} \quad \chi=\pi^{3/2}M\xi^2, \quad \mathrm{and} \quad \Delta\theta = \xi/M. \label{GaussianRateWidth}
\end{equation}
Hence we see that in addition to directionality, the rate is enhanced by a factor $\chi$.

Let us finally explain what the  initial approximation $\Gamma\ll v_0/L$ means in terms of the system parameters. Using (\ref{GaussianRateWidth}), this can be written as
\begin{equation}
\frac{\Omega^2}{\omega_0^2}\ll\frac{X_0/d_0}{M^2\pi^2\xi^2},
\end{equation}
which can be always satisfied by a proper tuning of the two--photon Rabi frequency.

This analysis shows that our scheme can be used to observe directional, superradiant emission by starting with a superfluid phase and tuning the laser wave vector so that $k_L=k_0$.

\section{Considerations about dimensionality \label{Dimensionality}}

So far we have allowed atoms in state $\left\vert b\right\rangle$ to move freely in 3D. However, the free motion of these atoms can be restricted to 2D or 1D by coupling them to an additional 1D or 2D deep optical lattice, respectively. Under these conditions the momentum of the free atoms has components only along the non-trapped directions, say $\mathbf{k}_{2D}=k_x\hat{\mathbf{x}}+k_y\hat{\mathbf{y}}$ and $\mathbf{k}_{1D}=k_x\hat{\mathbf{x}}$; from the point of view of the atoms in state $\left\vert a\right\rangle$, the dimensionality of the reservoir they are interacting through is therefore reduced to 2D or 1D. Hence, only sites laying in the same $z=const$ plane (for the 2D reservoir) or $(y,z)=const$ tube (for the 1D reservoir) would be connected through a two--point correlation function given by (\ref{Gjl}), but replacing $G(\tau)$ by the correlation functions
\begin{equation}
\fl G_{2\mathrm{D}}\left(\tau\right) = \frac{\Omega^2}{\left(1+\frac{\mathrm{i}}{2}\omega_0\tau\right)} \exp\left(i\Delta\tau\right) \quad \mathrm{and} \quad G_{1\mathrm{D}}\left(\tau\right) = \frac{\Omega^2}{\left(1+\frac{\mathrm{i}}{2}\omega_0\tau\right)^{1/2}} \exp\left(i\Delta\tau\right). \label{G21D}
\end{equation}
$G_{1\mathrm{D}}$ coincides with the correlation function of atoms embedded in isotropic band--gap materials \cite{John94,John95}, while $G_{2\mathrm{D}}$ has no quantum--optical analog to our knowledge.

Hence, with an additional trapping of the free atoms, our scheme could be useful to study low--dimensional reservoirs described by these correlation functions.

\section{Conclusions \label{Conclusions}}

To conclude, let us briefly summarize what we have shown in this article.

We have studied a model consisting of atoms trapped in an optical lattice (with trapping frequency $\omega_0$, lattice period $d_0$, and cubic geometry with $M^3$ sites) which are coherently driven to a non-trapped state through a Raman scheme having detuning $\Delta$ respect to the atomic transition, relative wave vector between the Raman lasers $\mathbf{k}_L$, and two--photon Rabi frequency $\Omega$. The limits of non-interacting bosons and hard--core bosons --where either one or zero atoms can be in a given lattice site-- have been considered for the trapped atoms, while interactions involving the free atoms have been neglected.

These are the main results that we have found (numbering follows the sections of the article):

\begin{itemize}

\item [(\ref{AtomsH})] We have first deduced the Hamiltonian of the system, showing that it is equivalent to that of a collection of harmonic oscillators or two--level systems for non-interacting atoms and hard--core bosons, respectively, interacting with a common radiation field consisting of massive particles. \vspace{6pt}

\item [(\ref{SingleDynamics})] We have then studied the emission of free atoms when the lattice sites emit independently. We have found a phase transition at $\tilde{\Delta}=\Delta-4\Omega^2/\omega_0=0$ from a regime ($\tilde{\Delta}<0$) where the trapped and free atoms form a bound state and there is a nonzero steady state population left in the lattice, to a radiative regime ($\tilde{\Delta}>0$) in which all the atoms leave the lattice eventually. We have identified the Markovian and non-Markovian regimes of the emission, showing that the condition for Markovianity is $\vert\tilde{\Delta}\vert\gg8\pi\Omega^4/\omega_0^3$.

    This behavior is analogous to the one predicted in \cite{John94,John95} for the spontaneous emission of atoms embedded in photonic band--gap materials. \vspace{6pt}

\item [(\ref{Collective})] As for collective effects, we have studied them by using a reduced master equation for the lattice within the Born--Markov approximation, explaining under which conditions this may hold in our system. This master equation has shown that the number of interacting sites is controlled by a single parameter, namely $\xi=1/d_0k_0$ with $k_0=X_0^{-1}\sqrt{2\vert\tilde{\Delta}\vert/\omega_0}$.

    Now let us summarize the main collective effects that we have predicted to appear.

    \begin{itemize}

     \item [(\ref{Extended})] For $\tilde{\Delta}<0$ and $\mathbf{k}_L=\mathbf{0}$, the evolution of the lattice follows an effective Hamiltonian: Extended hopping for non-interacting atoms and extended ferromagnetic interactions for hard--core bosons. The coupling between lattice sites has a Yukawa form, but approaches a Coulomb form as $\xi$ increases.

         This could be helpful for the simulation of condensed--matter phenomena requiring extended interactions. \vspace{6pt}

     \item [(\ref{Superradiance})] For $\tilde{\Delta}>0$ and $\mathbf{k}_L=\mathbf{0}$, the master equation has a collective dissipation term in addition to the Hamiltonian part. In the $\xi\ll1$ limit this dissipation term corresponds to the lattice sites emitting independently with the same rate, while in the $\xi\gg M$ limit it takes a Dicke-like form, and therefore superradiant effects are expected to appear. Hence, we have argued that the superradiant phase--transition could be observed in our system by varying the $\xi$ parameter.

         Indeed, we have shown that both atomic and harmonic oscillator superradiance can be observed: \vspace{6pt}

        \begin{itemize}

         \item [(\ref{AtomicSuper})] When starting with a Mott phase having one atom per site in the collisional blockade regime (which is the equivalent of a collection of excited atoms), there exists a critical value of $\xi$ above which the emission rate is enhanced at the initial stage of the emission, and its maximum occurs for $t\neq0$.

             This delay of the maximum rate was suggested by Dicke in his seminal work \cite{Dicke54}, and was shown to be a signature of the superradiant phase--transition for atomic samples \cite{Ernst68,Agarwal70,Rehler71}. \vspace{6pt}

         \item [(\ref{HarmonicSuper})] If the lattice atoms are non-interacting, there are two equivalent ways of observing superradiant behavior. One can start with a superfluid phase, and measure the time needed to radiate the lattice atoms, which should be reduced by a factor $M^3$ when going from the $\xi\ll1$ to the $\xi\gg M$ limit. Alternatively, one can start with a Mott phase having $N$ atoms in the lattice, and measure the number of trapped atoms left in the steady state; this should go from zero in the $\xi\ll1$ limit, to $N-N/M^3$ in the $\xi\gg M$ limit.

             This behavior is consistent with the results found in \cite{Agarwal70} for the spontaneous emission of a collection of harmonic oscillators, and hence our scheme provides a physical realization of this system. \vspace{6pt}

        \end{itemize}

    \item [(\ref{Directionality})] The last phenomenon linked to quantum--optics that we have shown is that superradiant, directional emission can be obtained in our system. In particular, we have proved that by starting from a superfluid phase distributed over enough lattice sites ($M\gg1$), matching $k_L$ to $k_0$, and working in the $\pi\xi>1$ regime, the atoms are emitted mostly within the $\mathbf{k}_L$ direction. If the condition $\xi/M\ll1$ is also satisfied, the beam has been shown to have angular width $\Delta\theta=\xi/M$ and a rate enhanced by a factor $\chi=\pi^{3/2}M\xi^2$.

        These results are in correspondence with those of \cite{Porras07a,Porras07}, where the emission of photons by regular arrays of atoms was studied.

    \end{itemize}

\item [(\ref{Dimensionality})] We have finally discussed what would change if the motion of the free atoms is restricted to 2D or 1D by using an additional optical lattice. We have shown that in this conditions the correlation function of the reservoir through which the lattice sites are interacting takes the form $G(\tau)\propto(1+\mathrm{i}\omega_0\tau/2)^{d/2}$, where $d$ is the dimensionality ($d=1$ and $2$ for 1D and 2D, respectively). For 1D this corresponds to the correlation function of the radiation field in an isotropic photonic band--gap material \cite{John94,John95}, while for 2D it has no quantum--optical counterpart.

\end{itemize}

All the proposed scenarios can be reached with currently available technology, and hence we believe that this work provides a solid base for considering optical lattices as quantum simulators of quantum--optical systems.

\ack

We would like to thank Miguel Aguado, Mari Carmen Ba\~{n}uls, Stephan D\"urr, G\'eza Giedke, Eugenio Rold\'an, and Germ\'an J. de Valc\'arcel for helpful discussions. CN-B would also like to thank Alberto Aparici for giving him access to \textit{Lux}, and the Theory Group at the Max-Planck-Institut f\"ur Quantenoptik as well as Profs. Susana Huelga and Martin Plenio at Ulm University for their hospitality. IdV acknowledges support and encouragement from Profs. Susana Huelga and Martin Plenio.

CN-B is a grant holder of the FPU program of the Ministerio de Ciencia e Innovaci\'on (Spain), and acknowledges support from the Spanish Government and the European Union FEDER through Project FIS2008-06024-C03-01. IdV has been supported by the EC under the grant agreement CORNER (FP7-ICT-213681). DP acknowledges contract RyC Y200200074, and spanish projects QUITEMAD and MICINN FIS2009-10061. JIC acknowledges support from the EU (AQUTE) and the DFG (FG 631).

\appendix

\section{Extracting information from the non-Markovian equation
(\ref{NonMarkovianA}) \label{Appendix I}}

In this Appendix we explain how to manipulate (\ref{NonMarkovianA}) by using Laplace transform techniques in order to understand the emission
properties of the system in the limit of independent lattice sites.

Making the Laplace transform of (\ref{NonMarkovianA}) the solution for the amplitude $A$ in Laplace space is found as
\footnote{
We define the Laplace transform of a function $f\left(t\right)$ as \cite{Riley06book}
\begin{equation}
\tilde{f}\left(s\right) = \mathcal{L}_s \left[f\left(t\right)\right] = \int_0^{+\infty} \mathrm{d}t\, \mathrm{e}^{-st} f\left(t\right),
\end{equation}
which satisfies the properties \cite{Riley06book}
\begin{equation}
\fl \mathcal{L}_s\left[\dot{f}\left(t\right)\right] = s \tilde{f}\left(s\right) - f\left(0\right) \quad \mathrm{and} \quad \mathcal{L}_s\left[\int_0^t \mathrm{d}t^\prime\, g\left(t-t^\prime\right) f\left(t^\prime\right)\right] = \tilde{g}\left(s\right) \tilde{f}\left(s\right).
\end{equation}
}
\begin{equation}
\tilde{A}\left(s\right) = \frac{1}{s+\tilde{G}\left(s\right)}. \label{Atrans}
\end{equation}
On the other hand, the Laplace transform of the correlation function (\ref{G}) is
\begin{equation}
\fl \tilde{G}\left(s\right) = \frac{4\Omega^2}{\omega_0} \left\{-\mathrm{i}+\mathrm{e}^{-\frac{2\left(\Delta+\mathrm{i}s\right)}{\omega_0}} \sqrt{\frac{2\pi\left(\Delta+\mathrm{i}s\right)}{\omega_0}} \left[1+\mathrm{erf}\left(\mathrm{i}\sqrt{\frac{2\left(\Delta+\mathrm{i}s\right)}{\omega_0}}\right)\right]\right\}, \label{Gtrans}
\end{equation}
or to first order in $\left(\Delta+\mathrm{i}s\right)/\omega_0$
\begin{equation}
\tilde{G}_\infty\left(s\right) = -\frac{4\mathrm{i}\Omega^2}{\omega_0} + \alpha\left(1+\mathrm{i}\right) \sqrt{2\pi\left(s-\mathrm{i}\Delta\right)}. \label{GtransStrong}
\end{equation}
Note that as we assume that $s/\omega_{0}\ll1$, this transform cannot describe properly times below $\omega_0^{-1}$, and this is why sometimes this limit receives the name `strong confinement', as it needs $\omega_0\rightarrow\infty$ or otherwise the initial steps of the evolution are lost in the description. Unlike in the general case where the transform of the correlation function has a really complicated form (\ref{Gtrans}), in this limit it is possible to make the inverse Laplace transform of $\tilde{A}\left(s\right)$. Hence, in the following we consider the two cases separately, as the
first case requires special attention in order to extract results.

\subsection{Evolution of the population in the strong confinement limit \label{AI evo}}

\begin{figure}[t]

\flushright \includegraphics[width=3.5in]{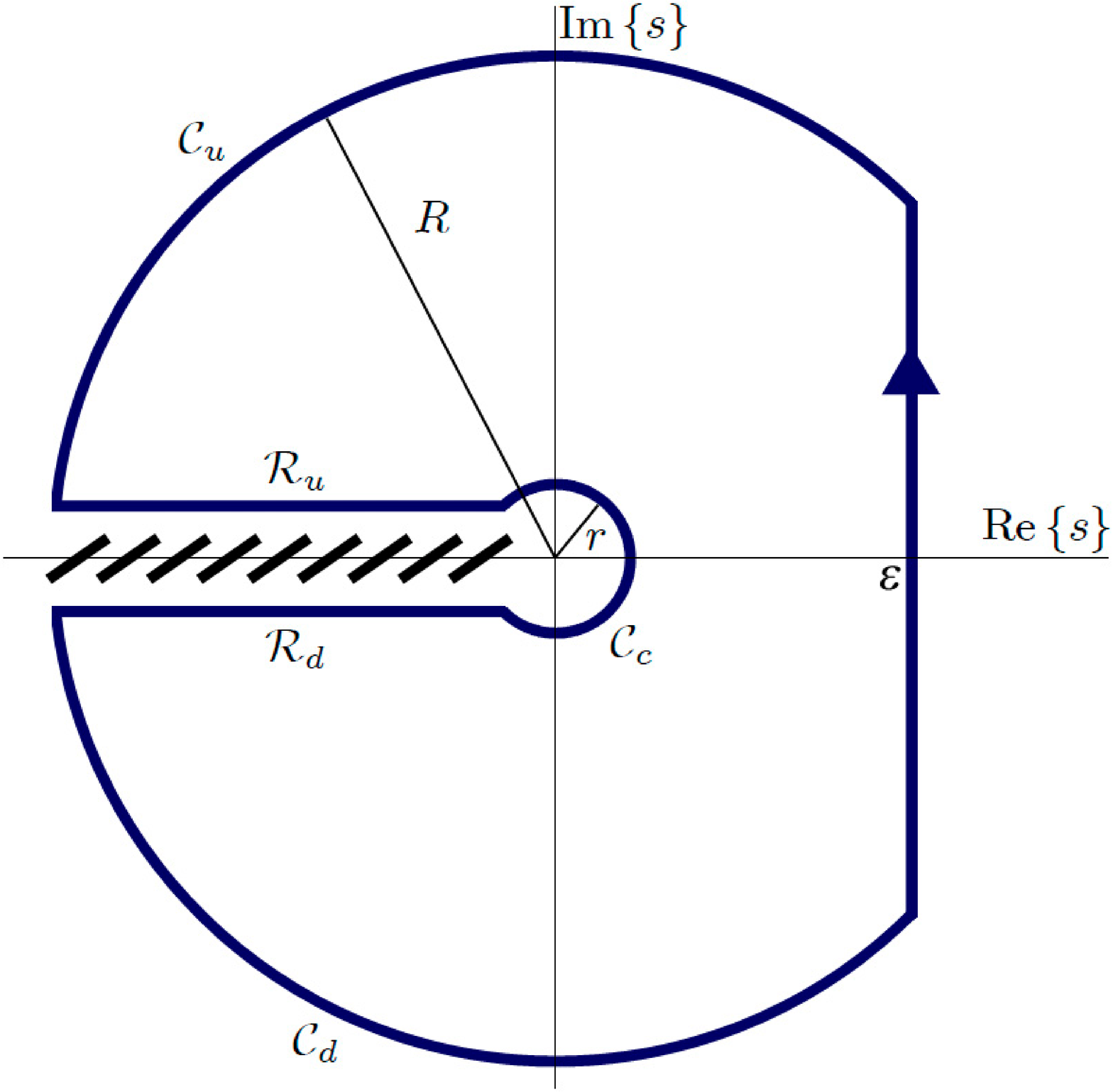}

\caption{Integration circuit in the complex--$s$ plane for the evaluation of the inverse Laplace transform.}

\label{Fig6}

\end{figure}

The Laplace transform of $A$ can be inverted as \cite{Riley06book}
\begin{equation}
\fl \mathrm{e}^{-\mathrm{i}\Delta t} A\left(t\right) = \frac{1}{2\pi\mathrm{i}}\int_{\epsilon-\mathrm{i}\infty}^{\epsilon+\mathrm{i}\infty} \mathrm{d}s\, \mathrm{e}^{st} \tilde{A}\left(s+\mathrm{i}\Delta\right) = \frac{1}{2\pi \mathrm{i}} \int_{\epsilon-i\infty}^{\epsilon+i\infty} \mathrm{d}s\, \frac{\mathrm{e}^{st}}{s+\mathrm{i}\tilde{\Delta}+\alpha\left(1+\mathrm{i}\right)\sqrt{2\pi s}}, \label{OurInverseLaplace}
\end{equation}
where we have used the `strong confinement' form of the correlation transform (\ref{GtransStrong}), and $\epsilon>0$ must be chosen so that all the poles of $\tilde{A}\left(s+\mathrm{i}\Delta\right)$ stay to the left of the $\mathrm{Re}\left\{s\right\}=\epsilon$ line. The multivaluated character of $\sqrt{s}$ forces us to define a branch cut in the complex-$s$ space; we chose to remove the branch $\arg\left\{s\right\}=\pi$, hence defining the phase of $s$ in the domain $\left]-\pi,\pi\right[$. Then, choosing the integration contour as that in figure \ref{Fig6}, we can write the needed integral as
\begin{equation}
\fl \int_{\epsilon-i\infty}^{\epsilon+i\infty} \mathrm{d}s K\left(s\right) = 2\pi \mathrm{i} \sum_j R_j - \lim_{\begin{array}{c} \scriptstyle r\rightarrow0 \\ \scriptstyle R\rightarrow\infty \end{array}} \left[\int_{\mathcal{R}_u} \mathrm{d}s K\left(s\right) + \int_{\mathcal{R}_d} \mathrm{d}s K\left(s\right)\right], \label{IntegralDecompos}
\end{equation}
being $R_j$ the residues of the kernel $K\left(s\right)=\tilde{A}\left(s+\mathrm{i}\Delta\right)\mathrm{exp}\left(st\right)$ at its poles \cite{Riley06book}. The integrals over the paths $\mathcal{C}_u$, $\mathcal{C}_d$, and $\mathcal{C}_c$ are zero when $r\rightarrow0$ and $R\rightarrow\infty$.

Making the variable change $s=\mathrm{e}^{\mathrm{i}\pi}x$ and $s=\mathrm{e}^{-\mathrm{i}\pi}x$ in the integrals along $\mathcal{R}_u$ and $\mathcal{R}_d$, respectively, we get
\begin{equation}
\fl \lim_{\begin{array}{c} \scriptstyle r\rightarrow0 \\ \scriptstyle R\rightarrow\infty \end{array}}\!\!\! \left[\int_{\mathcal{R}_u} \mathrm{d}s K\left(s\right) + \int_{\mathcal{R}_d} \mathrm{d}s K\left(s\right)\right] = 4\sqrt{\pi}\alpha \mathrm{e}^{-\mathrm{i}\pi/4}\int_0^{+\infty} \mathrm{d}x\, \frac{\sqrt{x}\mathrm{e}^{-xt}}{\left(-x+\mathrm{i}\tilde{\Delta}\right)^2+\mathrm{i}4\pi\alpha^2x}, \label{Integral}
\end{equation}
which is not analytical, but can be performed numerically very efficiently.

On the other hand, the poles of $K\left(s\right)$ are easily evaluated by making the change $s=\mathrm{i}x^{2}$, hence turning its
denominator into $x^2+2\sqrt{\pi}\alpha x+\tilde{\Delta}$ whose roots are
\begin{equation}
x_\pm = -\sqrt{\pi}\alpha \pm \sqrt{\pi\alpha^2-\tilde{\Delta}} = \left\vert x_\pm\right\vert \mathrm{e}^{\mathrm{i}\varphi_\pm};
\end{equation}
therefore the poles are simple and read $s_\pm=\mathrm{i}x_\pm^2$. In order to evaluate the residues, it is convenient to write the Kernel of
(\ref{OurInverseLaplace}) as
\begin{equation}
K\left(s\right) = \prod_{j=\pm} \frac{\sqrt{s}+\mathrm{e}^{\mathrm{i}\pi/4}x_j}{s-\mathrm{e}^{\mathrm{i}\pi/2}x_j^2} \mathrm{e}^{st},
\end{equation}
from which the residues are easily found as
\begin{equation}
R_+ = \frac {\sqrt{s_+} + \mathrm{e}^{\mathrm{i}\pi/4} x_+} {\sqrt{s_+} - \mathrm{e}^{\mathrm{i}\pi/4} x_-} \mathrm{e}^{s_+t} \quad \mathrm{and} \quad R_- = \frac {\sqrt{s_-} + \mathrm{e}^{\mathrm{i}\pi/4} x_-} {\sqrt{s_-} - \mathrm{e}^{\mathrm{i}\pi/4} x_+} \mathrm{e}^{s_-t}.
\end{equation}
The problem now is that the phase of $s_{\pm}$ must be defined in the interval $\left]-\pi,\pi\right[$ (remember the branch cut in figure \ref{Fig6}), and hence we have to analyze their phases as a function of the phases of $x_\pm$, which we take in the interval $\left[0,2\pi\right]$ as a convention. In terms of the parameters $\tilde{\Delta}$ and $\alpha$, we can define 3 regions:

\begin{itemize}

  \item In the region $\tilde{\Delta}<0$ we have $\varphi_+=0$ and $\varphi_-=\pi$. Hence, we have to write
\begin{equation}
s_+ = \mathrm{e}^{\mathrm{i}\pi/2}\left\vert x_+ \right\vert ^2 \quad \mathrm{and} \quad s_-=\mathrm{e}^{\mathrm{i}\pi/2}\left\vert x_-\right\vert^2 \mathrm{e}^{2\mathrm{i}\pi} \times \mathrm{e}^{-2\mathrm{i}\pi},
\end{equation}
that is, we have to decrease the phase of $s_-$ by $2\pi$ to keep it within the interval $\left]-\pi,\pi\right[$; with $s_\pm$ written in this way, the residues read
\begin{equation}
R_+=\frac{2\left\vert x_+\right\vert}{\left\vert x_+\right\vert + \left\vert x_-\right\vert} \mathrm{e}^{\mathrm{i}x_+^2t} \quad \mathrm{and} \quad R_-=0.
\end{equation}
\vspace{6pt}

  \item In the region $0<\tilde{\Delta}<\pi\alpha^2$, we have $\varphi_\pm=\pi$, and hence, we have to decrease the phase of both $s_\pm$ by $2\pi$ to keep them within the interval $\left]-\pi,\pi\right[$; therefore both residues are zero $R_\pm=0$. \vspace{6pt}

  \item In the region $\tilde{\Delta}>\pi\alpha^2$, we have $\varphi_+\in\left[\pi/2,\pi\right]$, and $\varphi_-=\left[\pi,3\pi/2\right]$. Hence, we need to write
\begin{equation}
s_\pm = \mathrm{e}^{\mathrm{i}\pi/2} \left\vert x_\pm\right\vert^2 \mathrm{e}^{2\mathrm{i}\varphi_\pm} \times \mathrm{e}^{-2\mathrm{i}\pi} \quad \mathrm{and} \quad s_-=\mathrm{e}^{\mathrm{i}\pi/2} \left\vert x_-\right\vert^2 \mathrm{e}^{2\mathrm{i}\varphi_-} \times \mathrm{e}^{-4\mathrm{i}\pi},
\end{equation}
that is, we decrease the phase of $s_+$ by $2\pi$ and the phase of $s_-$ by $4\pi$, so that they both stay in the interval $\left]-\pi,\pi\right[$, and obtain
\begin{equation}
R_+ = 0 \quad \mathrm{and} \quad R_-=\frac{2x_-}{x_--x_+}\mathrm{e}^{\mathrm{i}x_-^2t}.
\end{equation}

\end{itemize}

Introducing these results for the residues and the integral (\ref{Integral}) in (\ref{IntegralDecompos}), we arrive to the expression that we wanted to prove for the population $A\left(t\right)$, see (\ref{Asol}).

\subsection{Steady state population in the general case \label{AI ss}}

From the last discussion it should be clear that the long term population is dictated solely by the residues of $K\left(s\right)=\tilde{A}\left(s+\mathrm{i}\Delta\right)\mathrm{exp}\left(st\right)$. Moreover, only the residues corresponding to pure imaginary poles will remain in the $t\rightarrow\infty$ limit, and hence the steady state population will be given by
\begin{equation}
\left\vert A\left(t\rightarrow\infty\right)\right\vert^2 = \left\vert\sum_{s_j/\mathrm{Re}\left\{s_j\right\}=0} R_j\right\vert^2.
\end{equation}
Then, in order to know the steady state population in the finite trap case, we just need to find the pure imaginary poles of $K\left(s\right)$ with the transform of the correlation function given by (\ref{Gtrans}). By writing $s=\mathrm{i}x$ with $x\in\mathbb{R}$, this amounts for solving the equation
\begin{equation}
x + \tilde{\Delta} + 4\Omega^2 \exp\left(\frac{2x}{\omega_0}\right) \sqrt{\frac{2\pi x}{\omega_0^3}} \left[1-\mathrm{erf}\left(\sqrt{\frac{2x}{\omega_0}}\right)\right] = 0.
\end{equation}

This equation cannot be solved analytically, but it can be efficiently done numerically. On the other hand, once we know the solutions $x_j$ of this equation, we can use them to evaluate their associated residues as \cite{Riley06book}
\begin{equation}
R_j=\left\{ \mathrm{e}^{\mathrm{i}x_jt}\frac{\mathrm{d}}{\mathrm{d}s} \left[s+\mathrm{i}\Delta+\tilde{G}\left(s+\mathrm{i}\Delta\right)\right]_{s=\mathrm{i}x_j}\right\}^{-1}.
\end{equation}

This is the method that we have used to compute the results shown in figure \ref{Fig2}d.

\section{Evaluation of the Markov couplings (\ref{Couplings}) \label{Appendix II}}

In this appendix we show that in the strong confinement regime the Markov couplings have the form (\ref{Couplings}). After the Born--Markov approximation is carried, the effective coupling between different sites has the form
\footnote{
Note that $\varepsilon$ is more than just a mathematical artefact for regularizing the integral. It accounts for the decoherence channels not taken into account in the model, which make the correlation between lattice sites decay as $G_{\mathbf{j}-\mathbf{l}}(\tau)\exp(-\varepsilon\tau)$. Even if these channels appear at a time scale larger than the relevant times ($\varepsilon\ll1$), they remove the zeros of the kernel's denominator, showing that its singularities are not physical.
}
\begin{equation}
\Gamma_{\mathbf{j}-\mathbf{l}} = \int_0^\infty \mathrm{d}\tau\, G_{\mathbf{j}-\mathbf{l}}\left(\tau\right) = \lim_{\varepsilon\rightarrow0^+} \sum_\mathbf{k} \frac{g_k^2}{\hbar^2} \frac {\exp\left[\mathrm{i}\left(\mathbf{k}-\mathbf{k_L}\right)\cdot\mathbf{r}_{\mathbf{j}-\mathbf{l}}\right]} {\varepsilon+\mathrm{i}\Delta_k}.
\end{equation}
In the limit $k_L\ll X_0^{-1}$ (in which we can remove $\mathbf{k}_L$ from $g_k$ --see (\ref{Hpar})--), taking the continuous limit for the momentum, and performing the angular part of the integral over momenta by assuming that $\mathbf{r}_{\mathbf{j}-\mathbf{l}}$ is within the $z$--axis, this expression can be reduced to
\begin{eqnarray}
\Gamma_{\mathbf{j}-\mathbf{l}} = \frac{8X_0\Omega^2}{\mathrm{i}\pi^{1/2}\omega_0} &\mathrm{e}^{-\mathrm{i}\mathbf{k}_L\cdot\mathbf{r}_{\mathbf{j}-\mathbf{l}}} \left[\int_0^\infty \mathrm{d}k\, \mathrm{sinc}\left(kr_{\mathbf{j}-\mathbf{l}}\right) \exp\left(-X_0^2k^2\right)\right. \label{GjlTrans}
\\
&- \left. \lim_{\varepsilon\rightarrow0^+} \left(\tilde{\Delta}+\mathrm{i}\varepsilon\right) \int_0^\infty\mathrm{d}k\, \frac{\mathrm{sinc}\left(kr_{\mathbf{j}-\mathbf{l}}\right) \exp\left(-X_0^2k^2\right)}{\tilde{\Delta}+\mathrm{i}\varepsilon-\frac{\hbar k^2}{2m}}\right], \nonumber
\end{eqnarray}
The first integral is easily evaluated as
\begin{equation}
\int_0^\infty \mathrm{d}k\, \mathrm{sinc}\left(kr_{\mathbf{j}-\mathbf{l}}\right) \exp\left(-X_0^2k^2\right) = \frac{\pi}{2r_{\mathbf{j}-\mathbf{l}}} \mathrm{erf}\left(\frac{r_{\mathbf{j}-\mathbf{l}}}{2X_0}\right).
\end{equation}
The second integral is not analytical in general. Nevertheless, in the strong confinement regime both $\vert\tilde{\Delta}\vert/\omega_0$ and $X_0/d_0$ are small, and hence the Gaussian function in the Kernel can be approximated by one, so that the remaining integral reads
\begin{equation}
\fl \int_0^\infty\mathrm{d}k\, \frac{\mathrm{sinc}\left(kr_{\mathbf{j}-\mathbf{l}}\right)}{\tilde{\Delta}+\mathrm{i}\varepsilon-\frac{\hbar k^2}{2m}} = \frac{\pi}{2r_{\mathbf{j}-\mathbf{l}}(\tilde{\Delta}+\mathrm{i}\varepsilon)} \left[1-\exp\left(\mathrm{i}\frac{r_{\mathbf{j}-\mathbf{l}}}{X_0} \sqrt{\frac{2(\tilde{\Delta}+\mathrm{i}\varepsilon)}{\omega_0}}\right)\right],
\end{equation}
Introducing these integrals in (\ref{GjlTrans}) and talking the $\varepsilon\rightarrow0$ limit we arrive to expression (\ref{Couplings}) for the Markov couplings.

\section*{References}

\bibliography{BibForQOwithOL_bib}

\begin{thebibliography}{10}

\bibitem{Jaksch04rev}
Dieter {Jaksch} and Peter {Zoller}.
\newblock The cold atom {Hubbard} toolbox.
\newblock {\em Ann. Phys.}, 315:52--79, 2004.

\bibitem{Lewenstein07rev}
Maciej {Lewenstein}, Anna {Sanpera}, Verónica {Ahufinger}, Bogdan {Damski},
  Aditi {Send(De)}, and Ujjwal {Sen}.
\newblock Ultracold atomic gases in optical lattices: mimicking condensed
  matter physics and beyond.
\newblock {\em Adv. Phys.}, 56:243--379, 2007.

\bibitem{Bloch08rev}
Immanuel {Bloch}, Jean {Dalibard}, and Wilhelm {Zwerger}.
\newblock Many-body physics with ultracold gases.
\newblock {\em Rev. Mod. Phys.}, 80:885, 2008.

\bibitem{Jaksch98}
Dieter {Jaksch}, C~{Bruder}, Juan~Ignacio {Cirac}, Crispin~W. {Gardiner}, and
  Peter {Zoller}.
\newblock Cold bosonic atoms in optical lattices.
\newblock {\em Phys. Rev. Lett.}, 81:3108, 1998.

\bibitem{Greiner02}
Markus {Greiner}, Olaf {Mandel}, Tilman {Esslinger}, Theodor~W. {H\"ansch}, and
  Immanuel {Bloch}.
\newblock Quantum phase transition from a superfluid to a {Mott} insulator in a
  gas of ultracold atoms.
\newblock {\em Nature}, 415:39--44, 2002.

\bibitem{Schneider08}
U.~{Schneider}, L.~{Hackerm\"uller}, S.~{Will}, Th. {Best}, I.~{Bloch}, T.~A.
  {Costi}, R.~W. {Helmes}, D.~{Rasch}, and A.~{Rosch}.
\newblock Metallic and insulating phases of repulsively interacting fermions in
  a {3D} optical lattice.
\newblock {\em Science}, 322:1524, 2008.

\bibitem{Hadzibabic06}
Zoran {Hadzibabic}, Peter {Kr\"uger}, Marc {Cheneau}, Baptiste {Battelier}, and
  Jean {Dalibard}.
\newblock {Berezinskii–-Kosterlitz–-Thouless} crossover in a trapped atomic
  gas.
\newblock {\em Nature}, 441:1118--1121, 2006.

\bibitem{Anderlini07}
Marco {Anderlini}, Patricia~J. {Lee}, Benjamin~L. {Brown}, Jennifer
  {Sebby-Strabley}, William~D. {Phillips}, and J.~V. {Porto}.
\newblock Controlled exchange interaction between pairs of neutral atoms in an
  optical lattice.
\newblock {\em Nature}, 448:452--456, 2007.

\bibitem{deVega08}
In\'es {de Vega}, Diego {Porras}, and Juan~Ignacio {Cirac}.
\newblock Matter-wave emission in optical lattices: Single particle and
  collective effects.
\newblock {\em Phys. Rev. Lett.}, 101(26):260404, Dec 2008.

\bibitem{Dicke54}
R.~H. {Dicke}.
\newblock Coherence in spontaneous radiation processes.
\newblock {\em Phys. Rev.}, 93(1):99, 1954.

\bibitem{Gross82rev}
M.~{Gross} and S.~{Haroche}.
\newblock Superradiance: An essay on the theory of collective spontaneous
  emission.
\newblock {\em Physics Reports}, 93(5):301 -- 396, 1982.

\bibitem{Ernst68}
V.~Ernst and P.~Stehle.
\newblock Emission of radiation from a system of many excited atoms.
\newblock {\em Phys. Rev.}, 176(5):1456--1479, Dec 1968.

\bibitem{Agarwal70}
G.~S. Agarwal.
\newblock Master-equation approach to spontaneous emission.
\newblock {\em Phys. Rev. A}, 2(5):2038--2046, Nov 1970.

\bibitem{Rehler71}
Nicholas~E. Rehler and Joseph~H. Eberly.
\newblock Superradiance.
\newblock {\em Phys. Rev. A}, 3(5):1735--1751, May 1971.

\bibitem{Eberly06}
J.~H. Eberly.
\newblock Emission of one photon in an electric dipole transition of one among
  {$N$} atoms.
\newblock {\em Journal of Physics B: Atomic, Molecular and Optical Physics},
  39(15):S599, 2006.

\bibitem{Scully06}
Marlan~O. Scully, Edward~S. Fry, C.~H.~Raymond Ooi, and Krzysztof W\'odkiewicz.
\newblock Directed spontaneous emission from an extended ensemble of {$N$}
  atoms: Timing is everything.
\newblock {\em Phys. Rev. Lett.}, 96(1):010501, 2006.

\bibitem{Porras07a}
D.~Porras and J.~I. Cirac.
\newblock Quantum engineering of photon states with entangled atomic ensembles.
\newblock arXiv: 0704.0641, 2007.

\bibitem{Porras07}
D.~Porras and J.~I. Cirac.
\newblock Collective generation of quantum states of light by entangled atoms.
\newblock {\em Phys. Rev. A}, 78(5):053816, Nov 2008.

\bibitem{Scully09a}
Marlan~O. Scully.
\newblock Collective {Lamb} shift in single photon {Dicke} superradiance.
\newblock {\em Phys. Rev. Lett.}, 102(14):143601, 2009.

\bibitem{Scully09b}
Marlan~O. Scully and Anatoly~A. Svidzinsky.
\newblock The super of superradiance.
\newblock {\em Science}, 325(5947):1510--1511, 2009.

\bibitem{John94}
Sajeev John and Tran Quang.
\newblock Spontaneous emission near the edge of a photonic band gap.
\newblock {\em Phys. Rev. A}, 50(2):1764--1769, 1994.

\bibitem{John95}
Sajeev John and Tran Quang.
\newblock Localization of superradiance near a photonic band gap.
\newblock {\em Phys. Rev. Lett.}, 74(17):3419--3422, 1995.

\bibitem{John84}
Sajeev John.
\newblock Electromagnetic absorption in a disordered medium near a photon
  mobility edge.
\newblock {\em Phys. Rev. Lett.}, 53(22):2169--2172, 1984.

\bibitem{Chin10rev}
Cheng Chin, Rudolf Grimm, Paul Julienne, and Eite Tiesinga.
\newblock Feshbach resonances in ultracold gases.
\newblock {\em Rev. Mod. Phys.}, 82(2):1225--1286, 2010.

\bibitem{Paredes04}
Bel\'en Paredes, Artur Widera, Valentin Murg, Olaf Mandel, Simon Folling,
  Ignacio Cirac, Gora~V. Shlyapnikov, Theodor~W. H{\"a}nsch, and Immanuel
  Bloch.
\newblock {Tonks-Girardeau} gas of ultracold atoms in an optical lattice.
\newblock {\em Nature}, 429:277--281, 2004.

\bibitem{Greiner96book}
W.~Greiner and J.~Reinhardt.
\newblock {\em Field quantization}.
\newblock Springer Verlag, 1996.

\bibitem{Weisskopf30}
V.F. Weisskopf and E.~Wigner.
\newblock Berechnung der nat\"urlichen linienbreite auf grund der diracschen
  lichttheorie.
\newblock {\em Z. Phys.}, 63:54, 1930.

\bibitem{Scully97book}
Marlan~O. Scully and M.~Suhail Zubairy.
\newblock {\em Quantum optics}.
\newblock Cambridge University Press, 1997.

\bibitem{Breuer02book}
H.-P. Breuer and F.~Petruccione.
\newblock {\em The theory of open quantum systems}.
\newblock Oxford University Press, 2002.

\bibitem{Rousseau10}
V.~G. Rousseau, G.~G. Batrouni, D.~E. Sheehy, J.~Moreno, and M.~Jarrell.
\newblock Pure mott phases in confined ultracold atomic systems.
\newblock {\em Phys. Rev. Lett.}, 104(16):167201, 2010.

\bibitem{Friedberg72}
R.~Friedberg, S.~R. Hartmann, and J.~T. Manassah.
\newblock Limited superradiant damping of small samples.
\newblock {\em Physics Letters A}, 40(5):365 -- 366, 1972.

\bibitem{Friedberg74}
R.~Friedberg and S.~R. Hartmann.
\newblock Temporal evolution of superradiance in a small sphere.
\newblock {\em Phys. Rev. A}, 10(5):1728--1739, 1974.

\bibitem{Riley06book}
K.F. Riley, M.P. Hobson, and S.J. Bence.
\newblock {\em Mathematical methods for physics and engineering}.
\newblock Cambridge University Press, 2006.

\end{thebibliography}
\bibliographystyle{unsrt}

\end{document}